\documentclass[a4paper,11pt]{article}
\pdfoutput=1 

\usepackage{jcappub} 

\usepackage[T1]{fontenc} 
\usepackage{hyperref}
\usepackage{graphics}
\usepackage{amssymb, amsmath}

\usepackage{xcolor}

\author[1]{Rongpu Zhou}
\author[1,2]{Simone Ferraro}
\author[1,2,3]{Martin White}
\author[1]{Joseph DeRose}
\author[1,2]{Noah Sailer}
\author[1]{Jessica Aguilar}
\author[4]{Steven Ahlen}
\author[1]{Stephen Bailey}
\author[5]{David Brooks}
\author[1]{Todd Claybaugh}
\author[25]{Kyle Dawson}
\author[6]{Axel de la Macorra}
\author[7]{Biprateep Dey}
\author[5]{Peter Doel}
\author[12]{Andreu Font-Ribera}
\author[8,9]{Jaime E. Forero-Romero}
\author[1]{Satya Gontcho A Gontcho}
\author[1]{Julien Guy}
\author[1]{Anthony Kremin}
\author[1]{Andrew Lambert}
\author[26]{Laurent Le Guillou}
\author[1]{Michael Levi}
\author[10]{Christophe Magneville}
\author[11,12]{Marc Manera}
\author[13]{Aaron Meisner}
\author[14,12]{Ramon Miquel}
\author[15]{John Moustakas}
\author[27]{Adam~D.~Myers}
\author[7]{Jeffrey A. Newman}
\author[16]{Jundan Nie}
\author[17,18,19]{Will Percival}
\author[20]{Mehdi Rezaie}
\author[21]{Graziano Rossi}
\author[22]{Eusebio Sanchez}
\author[1]{David Schlegel}
\author[23]{Michael Schubnell}
\author[24]{Hee-Jong Seo}
\author[23]{Gregory Tarl\'{e}}
\author[16]{Zhimin Zhou}

\affiliation[1]{Lawrence Berkeley National Laboratory, 1 Cyclotron Road, Berkeley, CA 94720, USA}
\affiliation[2]{Department of Physics, University of California, Berkeley, CA 94720}
\affiliation[3]{Department of Astronomy, University of California, Berkeley, CA 94720}
\affiliation[4]{Physics Dept., Boston University, 590 Commonwealth Avenue, Boston, MA 02215, USA}
\affiliation[5]{Department of Physics \& Astronomy, University College London, Gower Street, London, WC1E 6BT, UK}
\affiliation[6]{Instituto de F\'{\i}sica, Universidad Nacional Aut\'{o}noma de M\'{e}xico,  Cd. de M\'{e}xico  C.P. 04510,  M\'{e}xico}
\affiliation[7]{Department of Physics \& Astronomy and Pittsburgh Particle Physics, Astrophysics, and Cosmology Center (PITT PACC), University of Pittsburgh, 3941 O'Hara Street, Pittsburgh, PA 15260, USA}
\affiliation[8]{Departamento de F\'isica, Universidad de los Andes, Cra. 1 No. 18A-10, Edificio Ip, CP 111711, Bogot\'a, Colombia}
\affiliation[9]{Observatorio Astron\'omico, Universidad de los Andes, Cra. 1 No. 18A-10, Edificio H, CP 111711 Bogot\'a, Colombia}
\affiliation[10]{IRFU, CEA, Universit\'{e} Paris-Saclay, F-91191 Gif-sur-Yvette, France}
\affiliation[11]{Departament de F\'{i}sica, Universitat Aut\`{o}noma de Barcelona, 08193 Bellaterra (Barcelona), Spain}
\affiliation[12]{Institut de F\'{i}sica d’Altes Energies (IFAE), The Barcelona Institute of Science and Technology, Campus UAB, 08193 Bellaterra Barcelona, Spain}
\affiliation[13]{NSF's NOIRLab, 950 N. Cherry Ave., Tucson, AZ 85719, USA}
\affiliation[14]{Instituci\'{o} Catalana de Recerca i Estudis Avan\c{c}ats, Passeig de Llu\'{\i}s Companys, 23, 08010 Barcelona, Spain}
\affiliation[15]{Department of Physics and Astronomy, Siena College, 515 Loudon Road, Loudonville, NY 12211, USA}
\affiliation[16]{National Astronomical Observatories, Chinese Academy of Sciences, A20 Datun Rd., Chaoyang District, Beijing, 100012, P.R. China}
\affiliation[17]{Waterloo Centre for Astrophysics, University of Waterloo, 200 University Ave W, Waterloo, ON N2L 3G1, Canada}
\affiliation[18]{Department of Physics and Astronomy, University of Waterloo, 200 University Ave W, Waterloo, ON N2L 3G1, Canada}
\affiliation[19]{Perimeter Institute for Theoretical Physics, 31 Caroline St. North, Waterloo, ON N2L 2Y5, Canada}
\affiliation[20]{Department of Physics, Kansas State University, 116 Cardwell Hall, Manhattan, KS 66506, USA}
\affiliation[21]{Department of Physics and Astronomy, Sejong University, Seoul, 143-747, Korea}
\affiliation[22]{CIEMAT, Avenida Complutense 40, E-28040 Madrid, Spain}
\affiliation[23]{Department of Physics, University of Michigan, Ann Arbor, MI 48109, USA}
\affiliation[24]{Department of Physics \& Astronomy, Ohio University, Athens, OH 45701, USA}
\affiliation[25]{Department of Physics and Astronomy, The University of Utah, 115 South 1400 East, Salt Lake City, UT 84112, USA}
\affiliation[26]{Sorbonne Universit\'{e}, CNRS/IN2P3, Laboratoire de Physique Nucl\'{e}aire et de Hautes Energies (LPNHE), FR-75005 Paris, France}
\affiliation[27]{Department of Physics and Astronomy, University of Wyoming, Laramie, WY 82071, USA}

\emailAdd{RongpuZhou@lbl.gov}
\emailAdd{sferraro@lbl.gov}
\emailAdd{mwhite@berkeley.edu}

\title{\boldmath DESI luminous red galaxy samples for cross-correlations}

\abstract{We present two galaxy samples, selected from DESI Legacy Imaging Surveys (LS) DR9, with approximately 20,000 square degrees of coverage and spectroscopic redshift distributions designed for cross-correlations such as with CMB lensing, galaxy lensing, and the Sunyaev-Zel'dovich effect. The first sample is identical to the DESI Luminous Red Galaxy (LRG) sample, and the second sample is an extended LRG sample with 2-3 times the DESI LRG density. We present the improved photometric redshifts, tomographic binning and their spectroscopic redshift distributions and imaging systematics weights, and magnification bias coefficients. The catalogs and related data products will be made publicly available. The cosmological constraints using this sample and Planck lensing maps are presented in a companion paper. We also make public the new set of general-purpose photometric redshifts trained using DESI spectroscopic redshifts, which are used in this work, for all galaxies in LS DR9.}

\arxivnumber{23MM.NNNNN}

\begin{document}
\maketitle
\flushbottom

\section{Introduction}
\label{sec:intro}

This decade will witness a dramatic increase in data available from large-scale structure (LSS) and cosmic microwave background (CMB) surveys, with measurements from the Dark Energy Survey (DES) \cite{DES:2016jjg}, the Dark Energy Spectroscopic Instrument (DESI) \cite{DESI,DESI_PART_II}, the Legacy Survey of Space and Time (LSST) on the Vera Rubin Observatory \cite{LSST}, Euclid \cite{EUCLID18}, the Atacama Cosmology Telescope (ACT) \cite{ACT}, the South Pole Telescope (SPT) \cite{Carlstrom:2009um} the Simons Observatory (SO) \cite{SO}, and others.
Each of these new facilities will enable transformative science, but with major leaps in capabilities occurring simultaneously with many probes, their combination could be even more powerful.
One of the most exciting scientific opportunities afforded by these surveys is the ability to cross-correlate the data sets that are overlapping on the sky.
The promise of cross-correlations is that they enable new science as well as increase the robustness of the core science being pursued by each project. Moreover, by breaking degeneracies between astrophysical and cosmological parameters, the overall statistical power can be greatly enhanced. The cross-correlation with samples in different redshift bins (tomography) can allow us to study the redshift dependence of the signal in question, and recover the time-evolution of various quantities.

In this paper, we define and characterize the photometric sample of DESI Luminous Red Galaxies \cite{desi_lrg_paper} (henceforth ``Main LRGs'') and a higher--density extended LRG sample (henceforth ``Extended LRGs''). 
We also present and publicly release a new set of general-purpose photo-$z$'s that uses DESI spectroscopic redshifts for training. 
We use the photo-$z$'s of the LRG samples to define four tomographic bins spanning the redshift range $0.4 \lesssim z \lesssim 1.1$. 
We also provide spectroscopic redshift distributions using DESI spectroscopic data and imaging systematics weights for each redshift bin, and we calculate the effect of lensing magnification on the galaxy surface densities. The high reliability of the redshift distributions (thanks to the very high spectroscopic success rate), the small fraction of contaminants (see table \ref{tab:summary}) and the high spatial uniformity make these samples very well suited to cross-correlations.

In particular, (an earlier version of) this Main LRG sample has been recently used in the tomographic analysis of Planck CMB lensing \cite{White:2021yvw}, for which this work serves as a companion paper; see appendix \ref{app:old_sample} for a discussion of the difference between this earlier sample and the updated LRG sample presented in this work. In \cite{White:2021yvw}, the authors used the galaxy auto-correlation, together with its cross-correlation with Planck CBM lensing to provide a ${\sim}\,4\%$ constraint on the amplitude of structure at low redshift with negligible uncertainties on the redshift distribution of galaxies, since it is directly measured from DESI spectroscopy. This follows previous work using previous releases of the same imaging data \cite{Hang21a,Hang21b,Kitanidis21}.

Other future applications include the cross-correlation with maps of the thermal or kinematic Sunyaev-Zel'dovich effect (tSZ and kSZ respectively), galaxy or CMB lensing measurements from current and future experiments, as well as maps of the Cosmic Infrared Background (CIB).

The remainder of the paper is organized as follows: in section \ref{sec:data} we describe the imaging data used to extract our sample. In section \ref{sec:selection}, we discuss the selection of the sample. In section \ref{sec:imaging_systematics} we explore the impact and mitigation of imaging systematics, while in section \ref{sec:slope} we measure the number count slope necessary to model magnification bias. We characterize the angular clustering of the samples and present forecasts for a few cross-correlation scenarios in section \ref{sec:clustering_and_forecasts}. We describe the data products in section \ref{sec:data_products}, and we conclude in section \ref{sec:conclusions}. Appendix \ref{app:old_sample} describes the difference between the Main LRG sample presented in this work and an earlier version used in \cite{White:2021yvw}.
Appendix \ref{app:photoz} contains a more detailed description of the new photometric redshifts. Appendix \ref{app:fibermagnitude} describes how we model the change in fiber-flux (which is used in the sample selections) for a magnified galaxy.

\section{Data}
\label{sec:data}

\subsection{DESI imaging data}
\label{sec:imaging}

The parent imaging is the DESI Legacy Imaging Surveys Data Release 9, which is a combination of optical and mid-infrared imaging used for DESI target selection. The data covers three optical bands ($g$, $r$, $z$), with four additional mid-IR bands ($W1-W4$) provided by the Wide-field Infrared Survey Explorer (WISE; \cite{Wright10}). Three individual surveys, using three telescopes, provide the optical imaging data: at the Cerro Tololo Inter-American Observatory is the Dark Energy Camera Legacy Survey (DECaLS; \cite{Dey19}) using the Blanco telescope, providing imaging in the South Galactic Cap (SGC) and the North Galactic Cap (NGC) at DEC<32.375$^{\circ}$, and at the Kitt Peak National Observatory are the Beijing–Arizona Sky Survey (BASS; \cite{bass_survey,Dey19}) using the Bok Telescope and the Mayall $z$-band Legacy Survey (MzLS; \cite{Dey19}) using the Mayall Telescope, providing imaging in the NGC at DEC>32.375$^{\circ}$. The DESI imaging also includes non-DECaLS observations from the DECam instrument, mainly from the Dark Energy Survey (DES; \cite{DES:2016jjg}). Hereafter, we denote the region with DECam imaging (i.e., DECaLS and DES) as Southern imaging (or simply South) and the region covered by BASS+MzLS as Northern imaging (or simply North).  These terms should not be confused with the NGC and SGC. There are small differences in the photometry between North and South, i.e., the same galaxy has slightly different magnitudes and colors in the North and South imaging. To account for these differences, we implement slightly different galaxy selection cuts and train the photo-$z$ algorithms separately in the two regions. Within the Southern imaging, the DES region is deeper than the DECaLS region, but we do not distinguish them in photo-$z$ estimation or sample selection because the effects of the different depths on the galaxy samples is relatively small and because the DECaLS/DES boundary is not well defined.

Images are processed and calibrated through the National Optical Astronomy Observatory (NOAO) Community Pipeline, then fed into \textit{The Tractor} \cite{Lang16a}, which uses forward-modeling to perform source extraction and produce probabilistic inference of source properties. Our analysis is based on Data Release 9 (DR9) of the DESI Legacy Imaging Surveys \cite{Dey19}, which is used for DESI target selection and contains 1,969,942,678 unique objects and covers 19,721 deg$^2$ in all three optical bands jointly. The BASS+MzLS imaging covers 5,064 deg$^2$ of the NGC with Declination$>32.375^{\circ}$, whereas DECam imaging covers the rest of the DR9 footprint. After applying the masks and footprint trimming described in section \ref{sec:masks}, the total area is roughly 16,700 deg$^2$, with BASS+MzLS (North), DECaLS and DES covering 4,200, 4,700 and 7,800 deg$^2$, respectively. The 5-$\sigma$ galaxy depths in magnitude (corrected for extinction; see definition in section \ref{sec:imaging_systematics}) for BASS+MzLS/DECaLS/DES are: 24.02/24.39/24.89 in $g$ band, 23.44/23.82/24.67 in $r$ band, and 22.97/22.95/23.45 in $z$ band. We will discuss in more detail the differences in the different imaging regions in section \ref{sec:imaging_systematics}.

Table \ref{tab:summary} lists the key information about the imaging coverage and the LRG samples.

\begin{table}
    \centering
    \begin{tabular}{ll}
    \hline
    Imaging coverage (with 1+ observations in $grz$ bands) & 19700 deg$^2$ \\
    Imaging coverage after applying masks and footprint trimming & 16700 deg$^2$ \\
    DESI spectroscopic coverage & 14800 deg$^2$ \\
    Surface density of Main LRGs & 600 deg$^{-2}$ \\
    Surface density of Extended LRGs & 1669 deg$^{-2}$ \\
    Comoving number density of Main LRGs & $5\times10^{-4}\ h^3\mathrm{Mpc}^{-3}$ \\
    Comoving number density of Extended LRGs & $\leq1.5\times10^{-3}\ h^3\mathrm{Mpc}^{-3}$ \\
    Stellar contamination in Main LRGs & 0.3\% \\
    Stellar contamination in Extended LRGs & 0.3\% \\
    Redshift failure rate of Main LRGs (in Main Survey) & 1.0\% \\
    Redshift failure rate of Extended LRGs (in Survey Validation) & 0.9\% \\
    \hline
    \end{tabular}
    \caption{\label{tab:summary} Key facts about the LRG samples. The second row is the ``clean'' footprint with the following cuts (see section \ref{sec:masks} for details): ``clean'' in the veto mask, 2+ observations in $grz$ bands, E(B-V)<0.15, and stellar density less than 2500 deg$^{-2}$; the surface densities in rows 4 and 5 are based on this area. ``Stellar contamination'' is the fraction of photometric LRGs (that are ``clean'' in the veto mask) that are classified as stars based on the DESI spectroscopic data. The last two columns list the fraction of spectroscopic LRGs that are rejected by redshift quality cuts (see section \ref{sec:specz}), and they are based on Main Survey data for the Main LRGs and Survey Validation data for the Extended LRGs.}
\end{table}

\subsubsection{Zero point correction at $\mathrm{Dec}<-29.25^{\circ}$}

Ref.~\cite{zhou_ls_vs_gaia} showed by comparing with synthetic photometry derived from Gaia XP spectra that in LS DR9 the photometric zero point has systematic offsets of up to 19 mmag at $\mathrm{Dec}<-29.25^{\circ}$ (where the calibration is based on internal calibration whereas Pan-STARRS \cite{chambers_panstarrs1_2016} is used at $\mathrm{Dec}>-29.25^{\circ}$) in $r$ and $z$ bands.
These offsets are large enough to affect the density of the LRG samples. 
We correct for these systematic zero point offsets by artificially modifying the fluxes (and inverse variances) of the objects for all objects at $\mathrm{Dec}<-29.25^{\circ}$. We model the smooth-varying offsets by fitting a monopole (for the average offset) and dipole (for the spatial variation) to the offset map from \cite{zhou_ls_vs_gaia} at $\mathrm{Dec}<-29.25^{\circ}$ using the healpy \cite{Zonca2019} \texttt{fit\_dipole} routine.

Specifically, the corrected magnitude for an object is $m_\mathrm{corr} = m_\mathrm{original} - \Delta m$, and the magnitude offset is $\Delta m = a + \textbf{v} \cdot \textbf{b}$, where $a$ and $\textbf{b}$ are the best-fit parameters, $\textbf{v} = (\cos \delta \cos \alpha, \cos \delta \sin \alpha, \sin \delta)$, and $\alpha$ and $\delta$ are the right ascension and declination, respectively. For the $r$ band, $a=-0.0054387$,  $\textbf{b}=(0.0025345, -0.0036355, 0.010219)$; for the $z$ band, $a=0.0144506$,  $\textbf{b}=(0.0000641, -0.013588, 0.0039126)$; the resulting $\Delta m$ is in units of magnitude. We visually verify that the dipole correction is sufficient to account for the large angular scale variations in the zero point offsets.

After obtaining the corrected photometric catalogs, we recompute the photo-$z$'s and rerun the LRG selection and photo-$z$ tomographic cuts for every object at $\mathrm{Dec}<-29.25^{\circ}$. We find that the zero point correction brings the surface densities consistent with the rest of the DES region. The changes in the surface density at $\mathrm{Dec}<-29.25^{\circ}$ compared to the original DR9 photometry are: -5.6\%, 0.5\%, 2.9\%, 9.4\% in bins 1-4 of the Main LRG subsamples, and -6.4\%, -0.1\%, 1.0\%, 7.3\% in bins 1-4 of the Extended LRG subsamples (the tomographic subsamples are described in section \ref{sec:bins}). The zero point correction also produces coherent changes in the photo-$z$ by as much as 0.01, and the exact change varies with redshift and location.

The LRG sample in \cite{White:2021yvw} was not corrected for the zero point offsets at $\mathrm{Dec}<-29.25^{\circ}$ (as we were not aware of this issue when it was published). We compare the angular power spectra of the sample in \cite{White:2021yvw} and the updated sample in appendix \ref{app:old_sample}.

\subsection{Spectroscopic redshifts}
\label{sec:specz}

We use spectroscopic data from the DESI Survey Validation \cite{desicollaboration_validation_2023} (henceforth ``SV1'') and the first year of DESI Main Survey (henceforth ``Y1'') to accurately characterize the redshift distributions and stellar contaminations. Specifically, we use the data reduction (the version internally named ``iron'') that will be published with the DESI Y1 Data Release. We also use the publicly available DESI redshifts in the photo-$z$ training, specifically the tile-based redshifts from the DESI Early Data Release \cite{desicollaboration_early_2023}.

A detailed characterization of the spectroscopic performance of the DESI LRGs is presented in \cite{desi_lrg_paper}. Here we briefly summarize the redshift performance with some numbers updated for the Y1 data. To remove incorrect redshifts, we apply the following quality cuts similar to those in \cite{desi_lrg_paper}: $\Delta \chi^2>15$, $z_\mathrm{redrock}<1.45$, and ZWARN=0. 
The fraction of objects that fail the quality cuts (``redshift failures'') is 1.0\% of the observed Main LRG targets, and the Extended LRGs (which have longer exposure times from SV1) have a similar failure rate. We estimate that less than 0.2\% of the objects that pass the quality cuts are catastrophic redshift failures (which we define as deviating by more than 1000 km/s from the true redshift) based on deep observations from SV1.
Only 0.3\% of the photometric LRGs are spectroscopically classified as stars. The failure rate for each redshift subsample is listed in tables \ref{tab:main_bins} and \ref{tab:extended_bins}. The note that the aforementioned redshift failure rates and stellar contamination rates are estimates for the ``clean'' sample with the veto masks (described in section \ref{sec:masks}) applied, and the rates are much higher for the masked objects.

The LRGs that are rejected by the redshift quality cuts are predominantly fainter objects, and therefore they are more likely to be at higher redshifts. To get unbiased redshift distributions, each galaxy is assigned a weight equal to the inverse of the predicted success rate. The success rate is predicted using the $z$-band fiber-magnitude and the effective exposure time; see section 4.4 of \cite{desi_lrg_paper} for details. For computing the redshift distributions, we also exclude 40 fibers that either have much higher than normal redshift failure rates or have anomalous redshift distributions; these issues are mostly caused by defects in the DESI spectrograph CCDs.

\subsection{Photometric redshifts}
\label{sec:photoz}

\begin{figure}
    \resizebox{0.5\columnwidth}{!}{\includegraphics{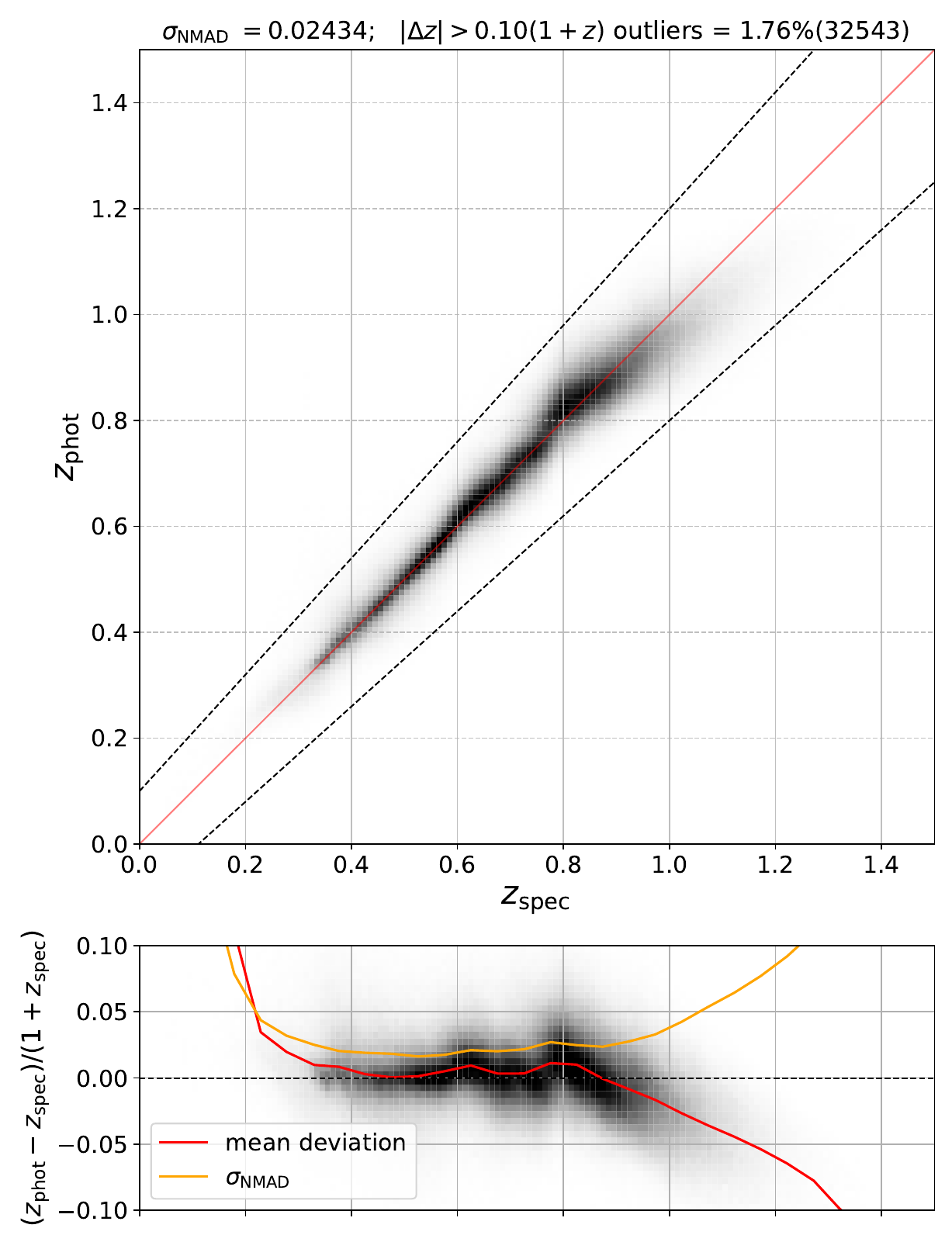}}
    \resizebox{0.5\columnwidth}{!}{\includegraphics{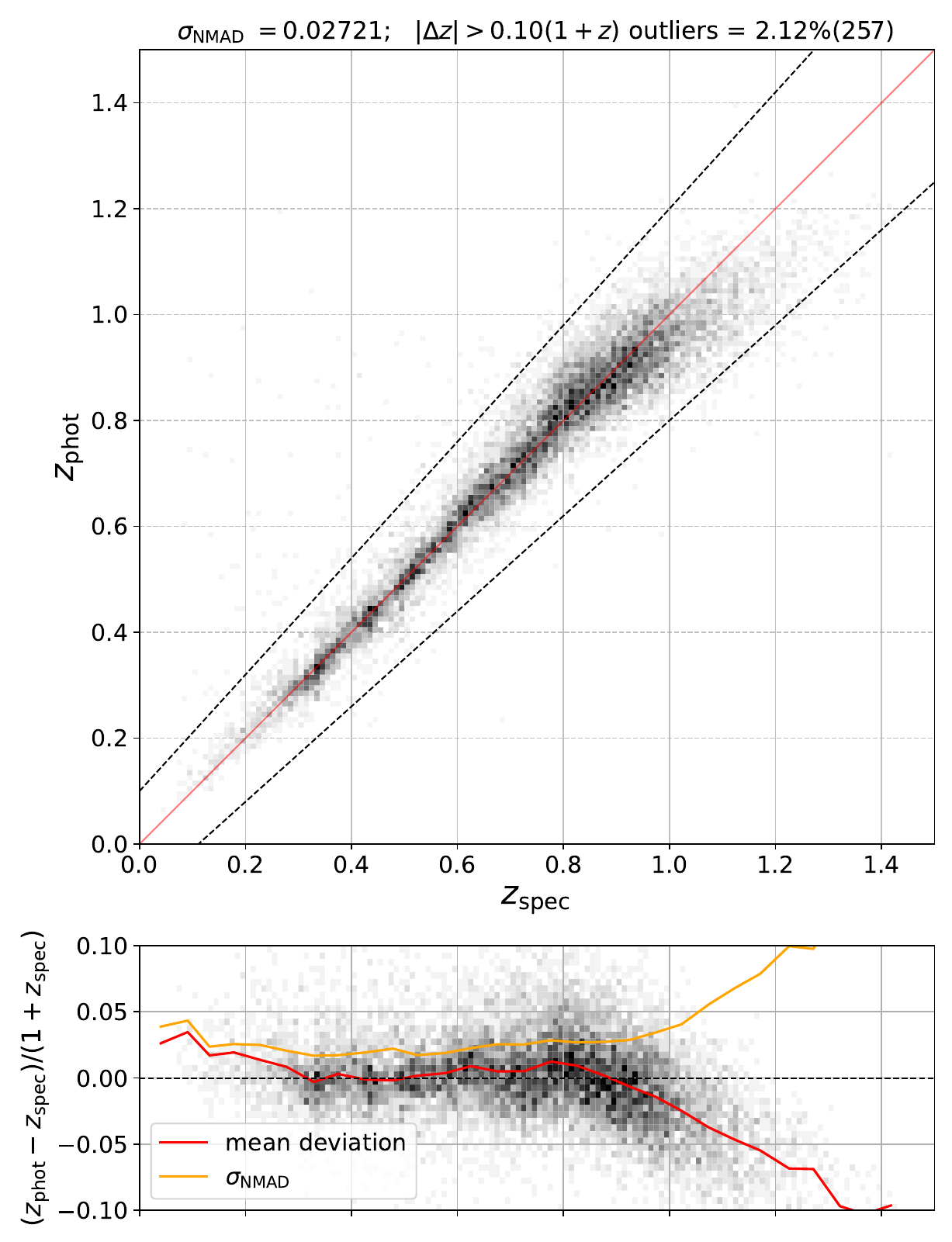}}
    \caption{Left panel: photo-$z$ vs spec-$z$ for 1.9 million galaxies in the Main LRG sample observed in DESI Y1. Right panel: same as the left panel but for 12k galaxies in the Extended LRG sample in DESI SV1. Both panels are based on photo-$z$'s from the Southern imaging. The photo-$z$ scatter is slightly larger in the (slightly shallower) Northern imaging.}
    \label{fig:main_lrg_pz_vs_sz_south}
\end{figure}

We provide a new set of photometric redshifts for all objects in LS DR9. While publicly available photo-$z$'s based on the method and training data described in \cite{Zhou++20} are already released as part of LS DR9, we compute this new set of photo-$z$'s to include improvements in the photo-$z$ algorithm and to take advantage of the newly available spectroscopic training data from the DESI survey.

Here we summarize the main changes from the previously released DR9 photo-$z$'s. We provide a more complete description of the photo-$z$ algorithm in Appendix \ref{app:photoz}. The first major change is that we have added DESI spectroscopic redshifts into the random forest training dataset. This spectroscopic sample includes galaxies from the Bright Galaxy Survey (BGS), Luminous Red Galaxies, Emission Line Galaxies (ELGs) and quasars observed during DESI Survey Validation (SV1) and the 1\% Survey (SV3). We also added SDSS DR16 quasars in the training data. This additional training data improves the overall photo-$z$ accuracy, especially in the North where training data was lacking, and reduces the bias at higher redshifts caused by the lack of spec-$z$'s coverage in that redshift-color-magnitude space. Some of the new training data (e.g., ELGs) does not significantly affect photo-$z$ performance of LRGs, but they improve the photo-$z$'s for other galaxy/quasar populations, and they are thus beneficial for this general-purpose photo-$z$ catalog. The second major change is that the photo-$z$'s are computed with 10-fold cross-validation, so that the photo-$z$ errors no longer depend on whether or not the object is in the spectroscopic training set. This change is less relevant for the tomographic analysis, but is important for analyses that utilize individual photo-$z$'s (e.g., galaxy-galaxy lensing).

We use DESI spectroscopic redshifts to assess the photo-$z$ performance of the Main and Extended LRG samples (described in section \ref{sec:selection}). Figure \ref{fig:main_lrg_pz_vs_sz_south} shows the spec-$z$ vs photo-$z$ for the Southern imaging. For the Main sample, the overall photo-$z$ error, quantified by the normalized median absolute deviation $\sigma_{\mathrm{NMAD}} \equiv 1.48 \times \mathrm{median}(|\Delta z|/(1 + z_{\mathrm{spec}}))$ where $\Delta z = z_{\mathrm{phot}} - z_{\mathrm{spec}}$, is 0.024 (0.026) in South (North), and the outlier fraction, defined as having $|\Delta z|>0.1 \times (1 + z_{\mathrm{spec}})$, is 1.8\% (2.0\%) in South (North). For the Extended sample, the photo-$z$ error is 0.027 (0.031) in South (North) and the outlier fraction is 2.1\% (3.2\%).

This new set of photo-$z$'s for all objects in DR9 is publicly available and described in section \ref{sec:data_products}. An earlier version of the photo-$z$'s was used for tomographic binning in \cite{White:2021yvw}, and we describe the differences in appendix \ref{app:old_sample}.

\section{Galaxy selection}
\label{sec:selection}

In this section, we present the two galaxy samples. The first sample is the DESI Main LRG sample, and the second sample is a higher-density galaxy sample based on the extended LRG selection in DESI Survey Validation. We also describe the tomographic binning based on photometric redshifts, and characterize the redshift distributions using DESI spectroscopic redshifts.

\subsection{Main LRG sample}

\begin{figure}
    \hspace*{-0.7cm}
    \resizebox{0.52\columnwidth}{!}{\includegraphics{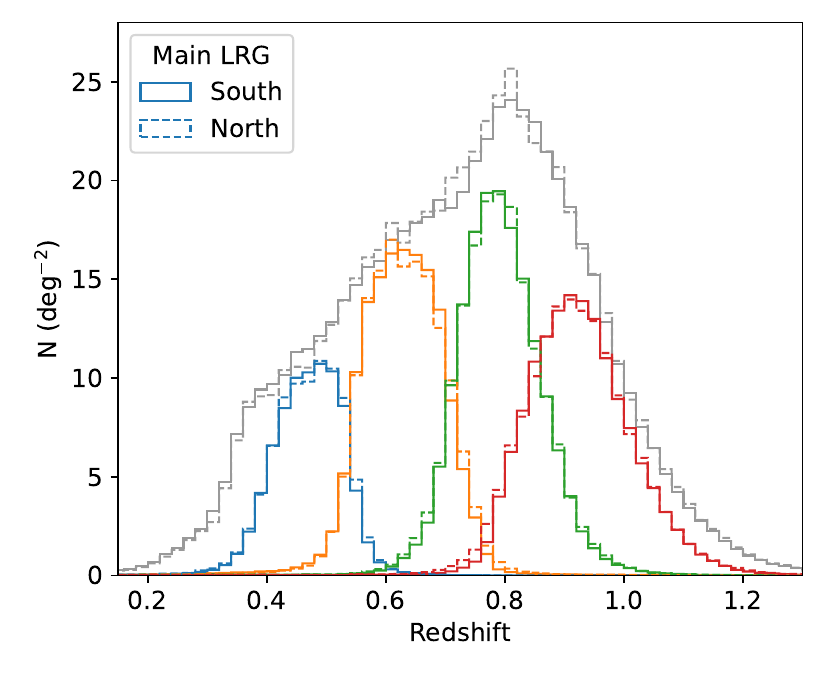}}
    \resizebox{0.52\columnwidth}{!}{\includegraphics{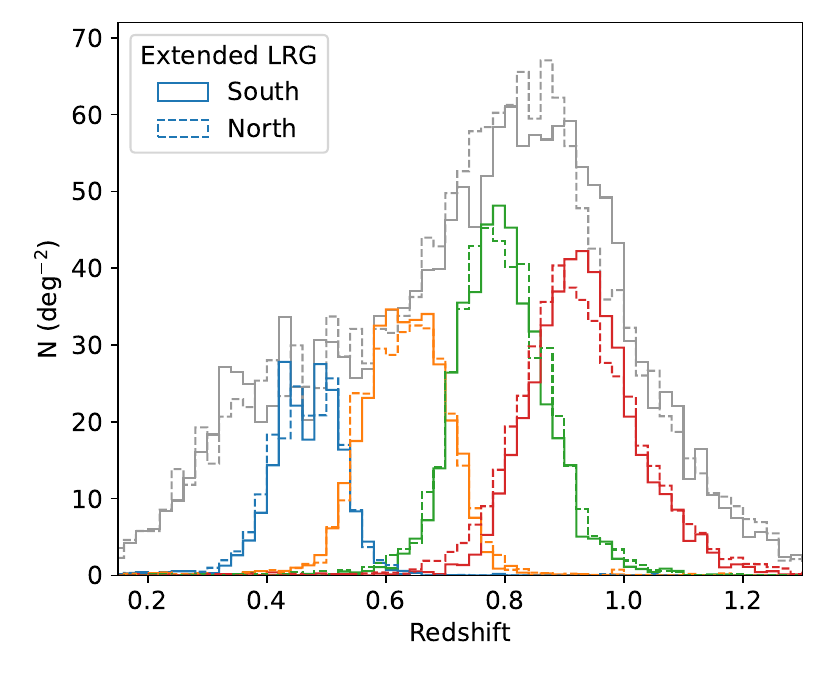}}
    \caption{Left panel: the redshift distributions of the full DESI LRG Main sample (in gray) and the four subsamples, derived directly from 2.3 million DESI spectroscopic redshifts. Right panel: same as the left panel but for the Extended LRG sample, based on 22k DESI spectroscopic redshifts. In both panels, units of the y-axis are the number of galaxies per square degree within the redshift bin with width $dz=0.02$.
    }
    \label{fig:dndz}
\end{figure}

\begin{figure}
    \hspace*{-0.7cm}
    \resizebox{0.52\columnwidth}{!}{\includegraphics{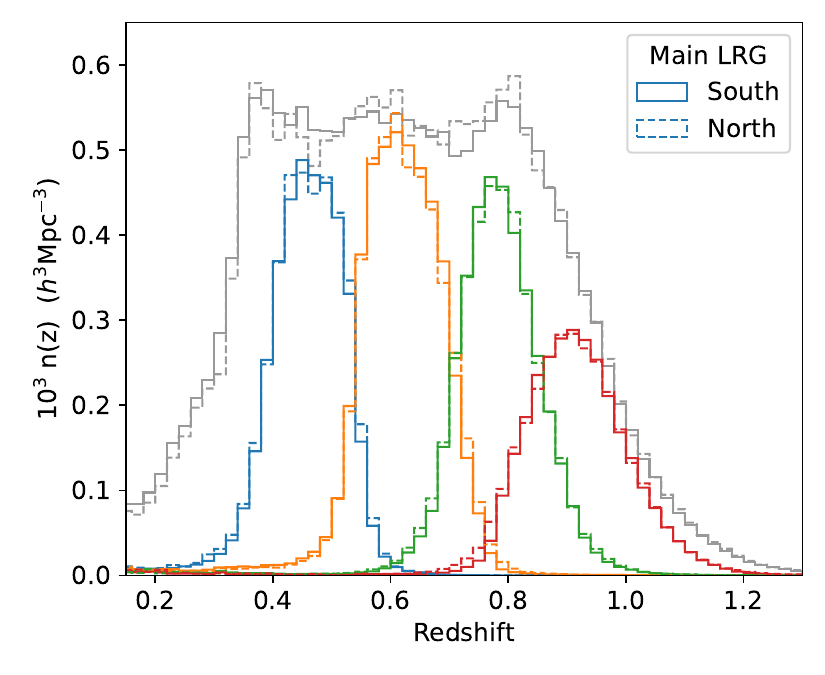}}
    \resizebox{0.52\columnwidth}{!}{\includegraphics{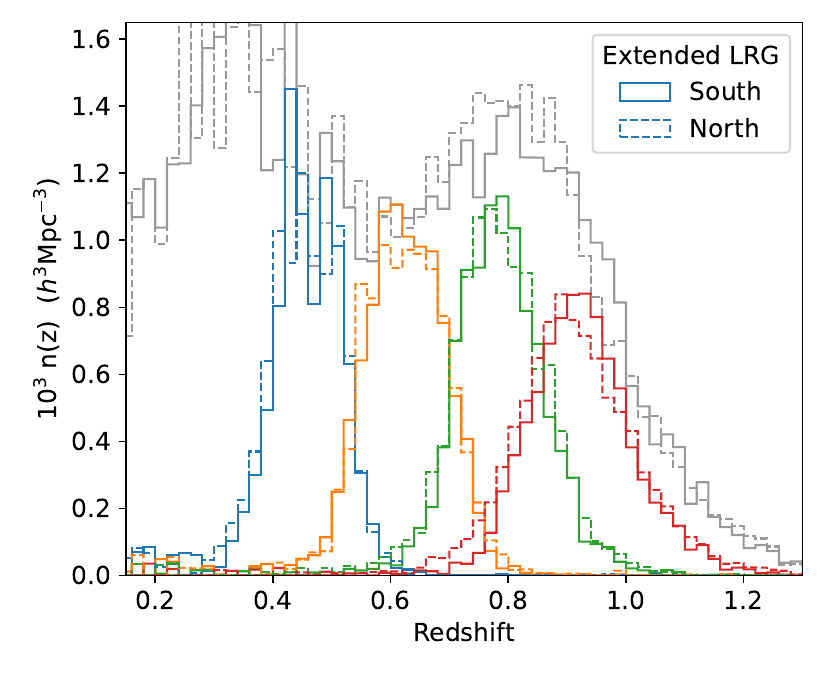}}
    \caption{Same as figure \ref{fig:dndz} but showing the comoving number densities as a function of redshift.
    }
    \label{fig:comoving_density}
\end{figure}

The Main LRG sample is identical to the DESI Main Survey LRG target sample described in \cite{desi_lrg_paper}. Here we briefly summarize the selection. The sample is selected with optical photometry in $g$, $r$ and $z$ bands and infrared photometry in WISE $W1$. A luminosity cut in $r-W1$ vs $W1$ produces a roughly constant comoving density of $5\times10^{-4}\ h^3\mathrm{Mpc}^{-3}$ up to $z=0.8$ and the faint limit in $z$-band fiber-magnitude produces a tail in the $N(z)$ that extends beyond $z=1.0$. A stellar rejection cut in $z-W1$ vs $r-z$ ensures very low (0.3\%) stellar contamination. And color cuts in $g-W1$ and $r-W1$ removes galaxies at $z<{\sim}\,0.4$.

The selection cuts for the Main LRG sample in the South:
\begin{subequations}
\begin{align}
    & z_\mathrm{fiber} < 21.60 \label{eq:mag-limit}\\
    & z-W1 > 0.8\times(r-z) - 0.6 \label{eq:non-stellar}\\
    & (g-W1>2.9) \ \mathrm{OR}\  (r-W1>1.8)  \label{eq:low-z}\\
   \begin{split}
    & ((r-W1 > 1.8\times(W1-17.14)) \ \mathrm{AND} \\
    & (r-W1 > W1-16.33)) \ \mathrm{OR}\ (r-W1>3.3) \label{eq:sliding-cut}
    \end{split}
\end{align}
\end{subequations}
and in the North:
\begin{subequations}
\begin{align}
    & z_\mathrm{fiber} < 21.61 \label{eq:mag-limit-north}\\
    & z-W1 > 0.8\times(r-z) - 0.6 \\
    & (g-W1>2.97) \ \mathrm{OR}\  (r-W1>1.8) \label{eq:low-z-north}\\
   \begin{split}
    & ((r-W1 > 1.83\times(W1-17.13)) \ \mathrm{AND} \\
    & (r-W1 > W1-16.31))\ \mathrm{OR}\ (r-W1>3.4)
    \end{split}
\end{align}
\end{subequations}
where $g$, $r$, $z$, $W1$ are extinction-corrected magnitudes and $z_\mathrm{fiber}$ is the $z$-band fiber-magnitude that corresponds to the flux within an aperture of diameter 1.5 arcsec (the size of the DESI fiber).

After applying the veto masks (see section \ref{sec:masks}), only 0.3\% of the photometric LRGs are stars. Stars are uncorrelated with extragalactic datasets and are therefore expected to give a negligible contribution to the cross-correlations (see for example section 7.1 of \cite{Krolewski:2019yrv} for a detailed discussion of this). Such a low stellar contamination fraction is also expected to give a negligible contribution to the galaxy auto-correlation, except possibly on the very largest scales. 
Since we'll mostly be working on scales that are not expected to be affected by this, we leave a full analysis of the effect of stellar contamination to future work.

We obtain the redshift distribution and comoving density of the Main LRG sample using the 2.3 million spectroscopic redshifts from the DESI Y1 data, and they are shown in figures \ref{fig:dndz} and \ref{fig:comoving_density}. The sample can be used as a single sample as just described, or in conjunction with the photometric redshifts provided. For tomographic analyses, we provide four subsamples selected with photometric redshifts, but whose redshift distributions have been measured with DESI spectroscopic redshifts. Their redshift distributions are thus known with very high precision and hence these are particularly well-suited to cross-correlation analyses. The tomographic subsamples are described in section \ref{sec:bins}.

\subsection{Extended LRG sample}

In addition to the DESI LRG sample, we created an ``Extended LRG sample'' with higher densities, useful for measurements that are limited by shot noise. 
While DESI will not observe the majority of the Extended LRGs, it has already obtained spectra for a large number of Extended LRGs from DESI SV1 which are sufficient for characterizing their redshift distributions.

The requirement of available spectra constrains the selection of the Extended LRG sample to within the DESI SV1 LRG selection. As described in appendix A of \cite{desi_lrg_paper}, the SV1 LRGs are selected with two sets of selection cuts, the ``optical'' selection and the ``IR'' selection (with a lower-density version of the SV1 IR selection ultimately being adopted as the Main LRG selection). The Extended LRG sample is a subset of the SV1 IR sample, with small offsets in the selection cuts compared to the SV1 IR selection to account for the differences in photometry between North and South. We use the same stellar-rejection cut as in the Main LRG selection (instead of the more relaxed version used in SV1 selections). And we only use fiber-magnitude for the faint limit (instead of also using the total magnitude in SV1 selections). The selection cuts for the Extended LRG sample in the South are:
\begin{subequations}
\begin{align}
    & z_\mathrm{fiber} < 21.96\\
    & z-W1 > 0.8\times(r-z) - 0.6\\
    & r-W1>1.0\\
    & (r-W1 > 1.8\times(W1-17.48)) \ \mathrm{OR}\ (r-W1>3.1)
\end{align}
\end{subequations}
and in the North:
\begin{subequations}
\begin{align}
    & z_\mathrm{fiber} < 22.0 \\
    & z-W1 > 0.8\times(r-z) - 0.6 \\
    & r-W1>1.03 \\
    & (r-W1 > 1.8\times(W1-17.44)) \ \mathrm{OR}\ (r-W1>3.2)
\end{align}
\end{subequations}

The SV1 data includes 34 DESI dark-time tiles covering roughly 230 deg$^2$ (130 deg$^2$ in the South and 100 deg$^2$ in the North), with roughly 6\% fiber-assignment completeness. We use 22k spectroscopic redshifts of the Extended LRGs from the DESI SV1 data to obtain their redshift distribution and comoving density, and they are shown in figures \ref{fig:dndz} and \ref{fig:comoving_density}. The $N(z)$ is noisier than that of the Main LRGs due to the smaller area and smaller number of spectroscopic redshifts available.

\subsection{Masks and imaging quality cuts}
\label{sec:masks}

The DESI LRG selection already implements some minimum masking using mask bits in DR9\footnote{\url{https://www.legacysurvey.org/dr9/bitmasks/\#maskbits}}: the BRIGHT, GALAXY and CLUSTER masks. A significant fraction of contaminants, mostly stars (especially at lower galactic latitudes), remain after the minimum masking. Therefore, in addition to these minimal target masks, we recommend applying the ``veto'' masks described in \cite{desi_lrg_paper} (see section 2.4 of that reference) that remove additional contaminated areas around bright stars as well as areas affected by imaging artifacts and unrelated astrophysical foregrounds (large galaxies, planetary nebulae, etc.).

Optionally, one could apply the additional imaging quality cuts that were used in \cite{White:2021yvw}:
1) Require two or more exposures in the $g$, $r$ and $z$ bands. 2) Require $E(B-V)<0.15$ (using the $E(B-V)$ map from \cite{schlegel_maps_1998}). 3) Apply a cut on stellar density: using the stellar density map provide by \cite{myers_targetselection_2023} (which is based on Gaia stars with $12\leq G < 17$) and degraded to a HEALPix resolution of nside=64, and requiring the stellar density in each pixel to be less than 2500 per deg$^2$. 

For the requirement of $2+$ exposures, instead of using the ``NOBS'' values in the DR9 tractor catalogs, we use the ``PIXEL\_NOBS'' values extracted from brick-level images (which are also used for randoms). This is because the tractor `NOBS'' values do not exactly match the ``NOBS'' values in the randoms\footnote{\url{https://www.legacysurvey.org/dr9/issues/\#nobs-differs-between-the-tractor-catalogs-and-random-catalogs}}, whereas the ``PIXEL\_NOBS'' values exactly match the survey geometry of the randoms.

\subsection{Tomographic bins}
\label{sec:bins}

\begin{table}[]
    \centering
    \begin{tabular}{c|cccccccc}
        Bin & North selection                   & South selection                     & $\bar{n}_\theta$ & $\bar{z}$ & $\Delta z$ & $f_\mathrm{star}$ & $f_\mathrm{fail}$ \\ \hline
         1  & $0.400\leq z_\mathrm{phot}<0.545$ & $0.400\leq z_\mathrm{phot}<0.540$   & 81.9             &  0.470    & 0.063      & 0.2\%             & 0.3\% \\
         2  & $0.545\leq z_\mathrm{phot}<0.719$ & $0.540\leq z_\mathrm{phot}<0.713$   & 148.1            &  0.628    & 0.074      & 0.1\%             & 0.7\%  \\
         3  & $0.719\leq z_\mathrm{phot}<0.851$ & $0.713\leq z_\mathrm{phot}<0.860$   & 162.4            &  0.791    & 0.078      & 0.2\%             & 1.0\%  \\
         4  & $0.851\leq z_\mathrm{phot}<1.024$ & $0.860\leq z_\mathrm{phot}<1.020$   & 148.3            &  0.924    & 0.096      & 0.4\%             & 1.3\% 
    \end{tabular}
    \caption{Key facts about the Main LRG subsamples. Here $ z_\mathrm{phot}$ is the \texttt{Z\_PHOT\_MEDIAN} value from the photo-$z$ catalog, $\bar{n}_\theta$ is the surface density in deg$^{-2}$, $\bar{z}$ is the mean redshift, $\Delta z$ is the standard deviation in $z$, $f_\mathrm{star}$ is the star fraction and $f_\mathrm{fail}$ is the redshift failure rate. The values of $\bar{z}$, $\Delta z$, $f_\mathrm{star}$ and $f_\mathrm{fail}$ are derived from the DESI Y1 data.}
    \label{tab:main_bins}
\end{table}

\begin{figure}
    \hspace*{-1.4cm}
    \resizebox{0.565\columnwidth}{!}{\includegraphics{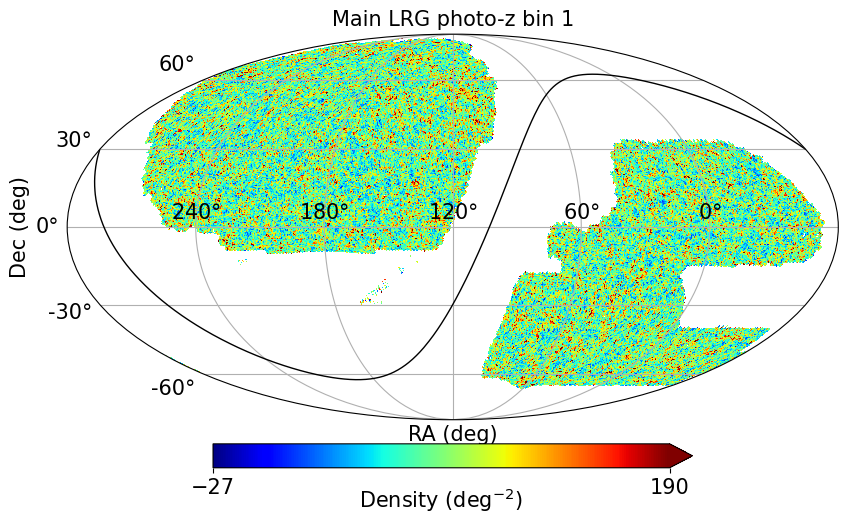}}
    \resizebox{0.565\columnwidth}{!}{\includegraphics{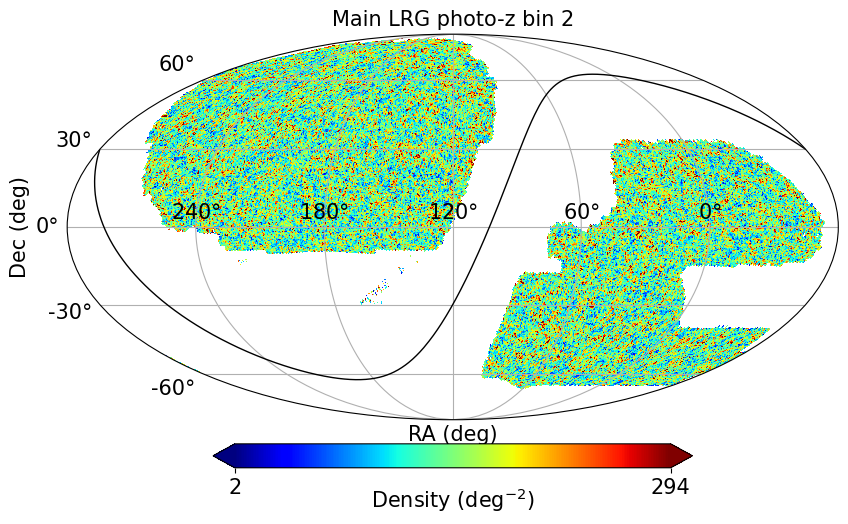}}
    \hspace*{-1.4cm}
    \resizebox{0.565\columnwidth}{!}{\includegraphics{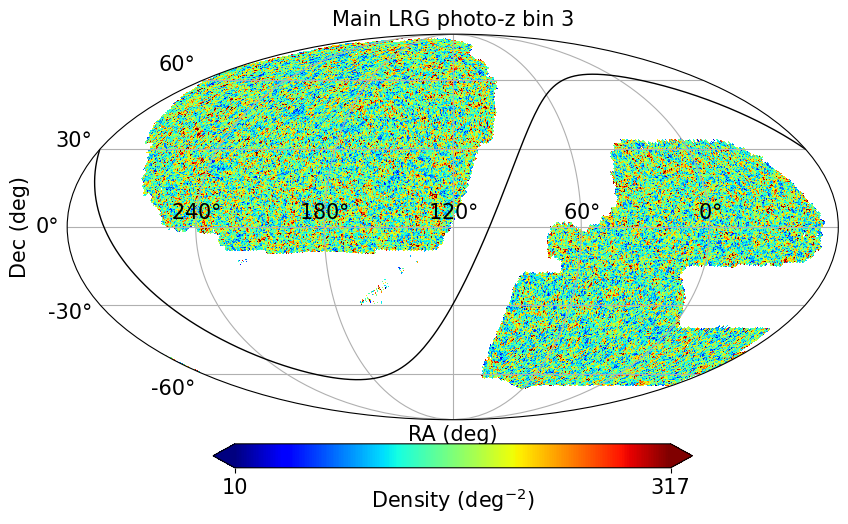}}
    \resizebox{0.565\columnwidth}{!}{\includegraphics{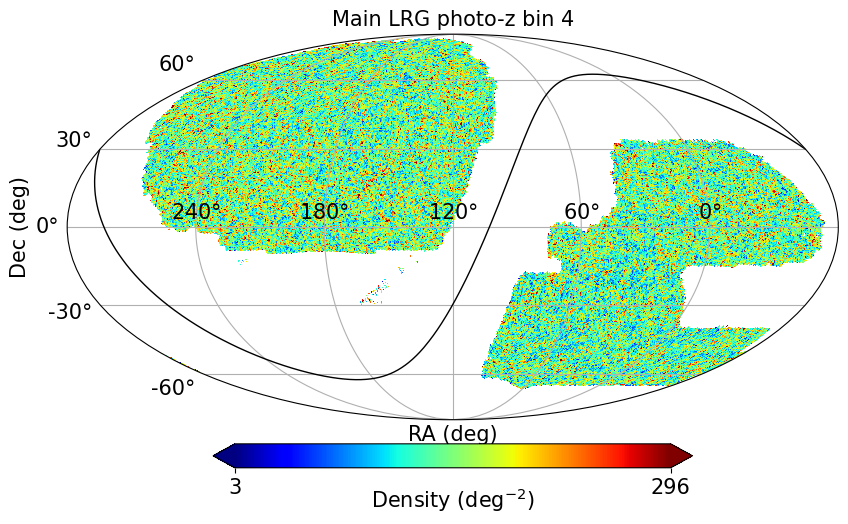}}
    \caption{Density maps of the Main LRG subsamples (after applying the veto masks and requiring 2+ observations in $grz$ bands) with HEALPix NSIDE=128. Note that additional cuts on the footprint described in section \ref{sec:masks} (that remove areas with high Galactic extinction and high stellar density) are not applied here, nor are the imaging weights in section \ref{sec:imaging_systematics} applied.}
    \label{fig:main_maps}
\end{figure}

\begin{figure}
    \hspace*{-1.4cm}
    \resizebox{0.565\columnwidth}{!}{\includegraphics{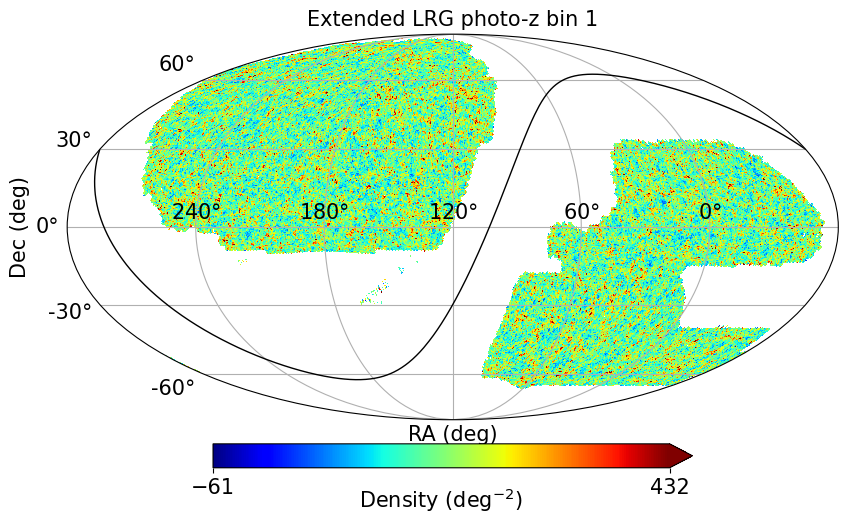}}
    \resizebox{0.565\columnwidth}{!}{\includegraphics{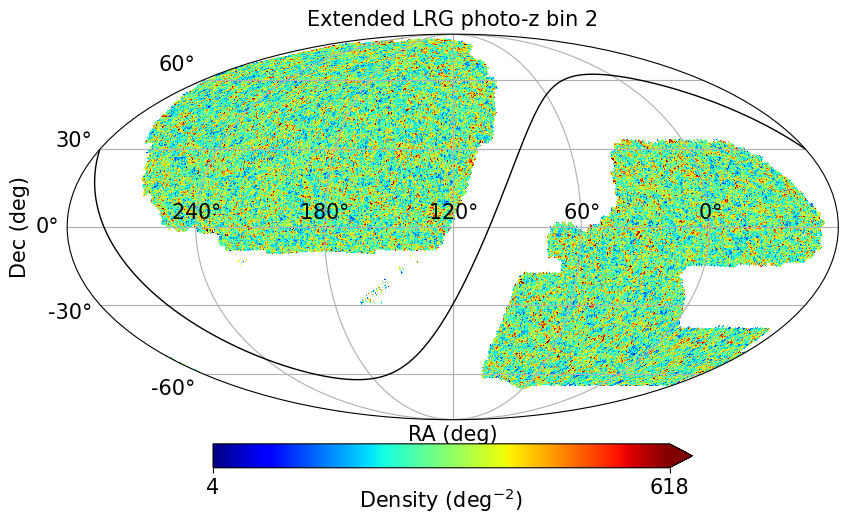}}
    \hspace*{-1.4cm}
    \resizebox{0.565\columnwidth}{!}{\includegraphics{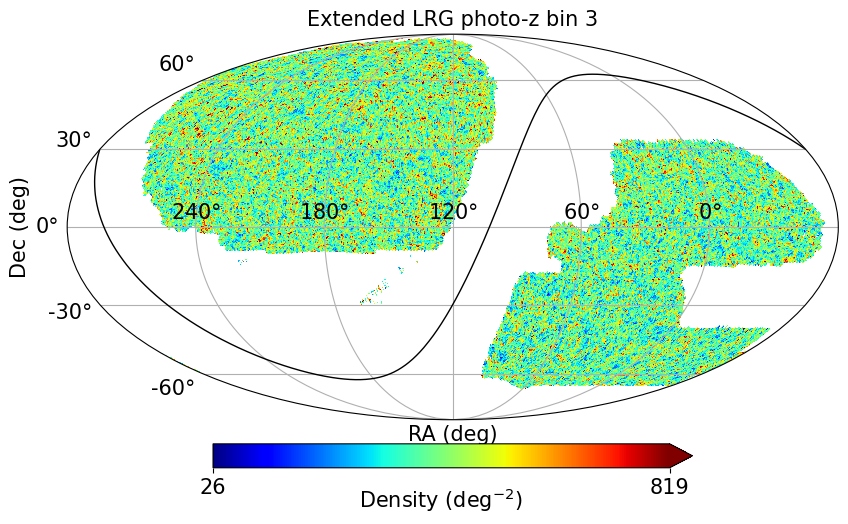}}
    \resizebox{0.565\columnwidth}{!}{\includegraphics{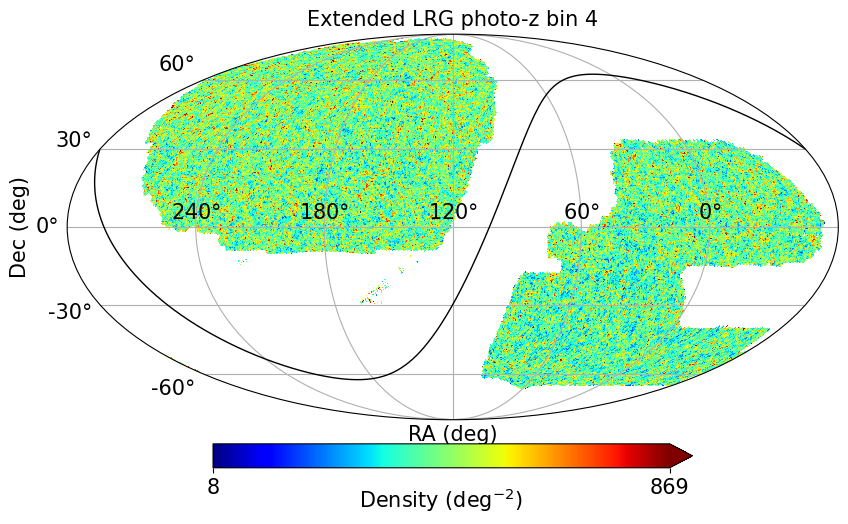}}
    \caption{As figure \ref{fig:main_maps}, but for the Extended LRG subsamples.}
    \label{fig:extended_maps}
\end{figure}

For some applications, e.g.\ the cross-correlation with CMB lensing, it is more useful to have subsamples of galaxies in tomographic bins with well-characterized redshift distributions, rather than individual galaxies with photometric redshifts. Here we divide each LRG sample into 4 bins using photometric redshifts, specifically, using the \texttt{Z\_PHOT\_MEDIAN} value from the photo-$z$ catalog. The photo-$z$ cuts and summary statistics of each subsample are listed in tables \ref{tab:main_bins} and \ref{tab:extended_bins}. We excluded objects at the tails of the $z_\mathrm{phot}$ distribution from the subsamples to achieve more compact redshift distributions for subsamples 1 and 4.
We measure the redshift distributions of the subsamples using spectroscopic redshifts from DESI. For tomographic analyses, the spectroscopic $N(z)$ should be preferred over the use of individual photometric redshifts. Density maps of the Main and Extended LRG subsamples are shown in figures \ref{fig:main_maps} and \ref{fig:extended_maps}.

\begin{table}[]
    \centering
    \begin{tabular}{c|cccccccc}
        Bin & North selection                   & South selection                     & $\bar{n}_\theta$ & $\bar{z}$ & $\Delta z$ & $f_\mathrm{star}$ & $f_\mathrm{fail}$ \\ \hline
         1  & $0.400\leq z_\mathrm{phot}<0.545$ & $0.400\leq z_\mathrm{phot}<0.540$   & 185.5            &  0.467    & 0.065      & 0.3\%             & 0.1\% \\
         2  & $0.545\leq z_\mathrm{phot}<0.719$ & $0.540\leq z_\mathrm{phot}<0.713$   & 311.0            &  0.633    & 0.077      & 0.2\%             & 0.3\%  \\
         3  & $0.719\leq z_\mathrm{phot}<0.854$ & $0.713\leq z_\mathrm{phot}<0.860$   & 422.6            &  0.794    & 0.086      & 0.2\%             & 0.7\%  \\
         4  & $0.854\leq z_\mathrm{phot}<1.010$ & $0.860\leq z_\mathrm{phot}<1.000$   & 438.4            &  0.929    & 0.102      & 0.4\%             & 0.9\% 
    \end{tabular}
    \caption{As table \ref{tab:main_bins} but for the extended LRG subsamples. The values of $\bar{z}$, $\Delta z$, $f_\mathrm{star}$ and $f_\mathrm{fail}$ are derived from the DESI SV1 data.}
    \label{tab:extended_bins}
\end{table}

\section{Imaging and foreground systematics}
\label{sec:imaging_systematics}

Here we study the impact of imaging and foreground systematics, such as variations in depth, seeing and Galactic extinction $E(B-V)$, on the density and redshift distribution of the binned LRG subsamples.

\begin{figure}
    \hspace*{-0.7cm}
    \resizebox{1.06\columnwidth}{!}{\includegraphics{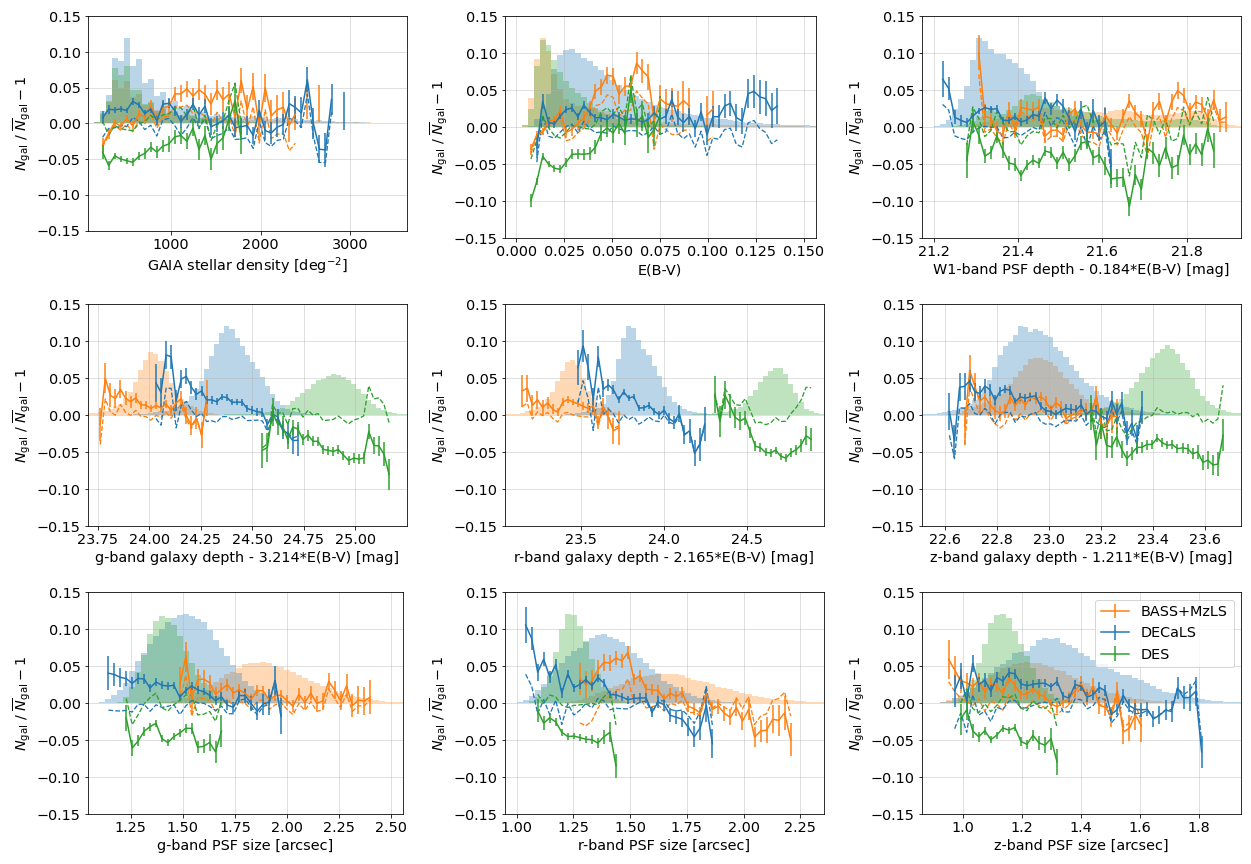}}
    \caption{Density trends with imaging/foreground systematics properties for the 4th redshift bin of the Main sample (based on HEALPix maps with NSIDE=256). The solid lines are $N_\mathrm{gal}/\overline{N}_\mathrm{gal}-1$, where $N_\mathrm{gal}$ is the average density within each bin in the systematics property, and $\overline{N}_\mathrm{gal}$ is the average number density over the full final footprint. The dashed lines are after applying the linear weights. The histograms show the normalized number of HEALPix pixels in each systematics bin for each region. The $E(B-V)$ term in the depth values is added to account for Galactic extinction. Note that while the trends with Gaia stellar density and $W1$ depth are shown here, they are not used as inputs for the weights (see section \ref{sec:imaging_systematics}).
    }
    \label{fig:main_lrg_density_trends}
\end{figure}

\begin{figure}
    \hspace*{-0.7cm}
    \resizebox{1.06\columnwidth}{!}{\includegraphics{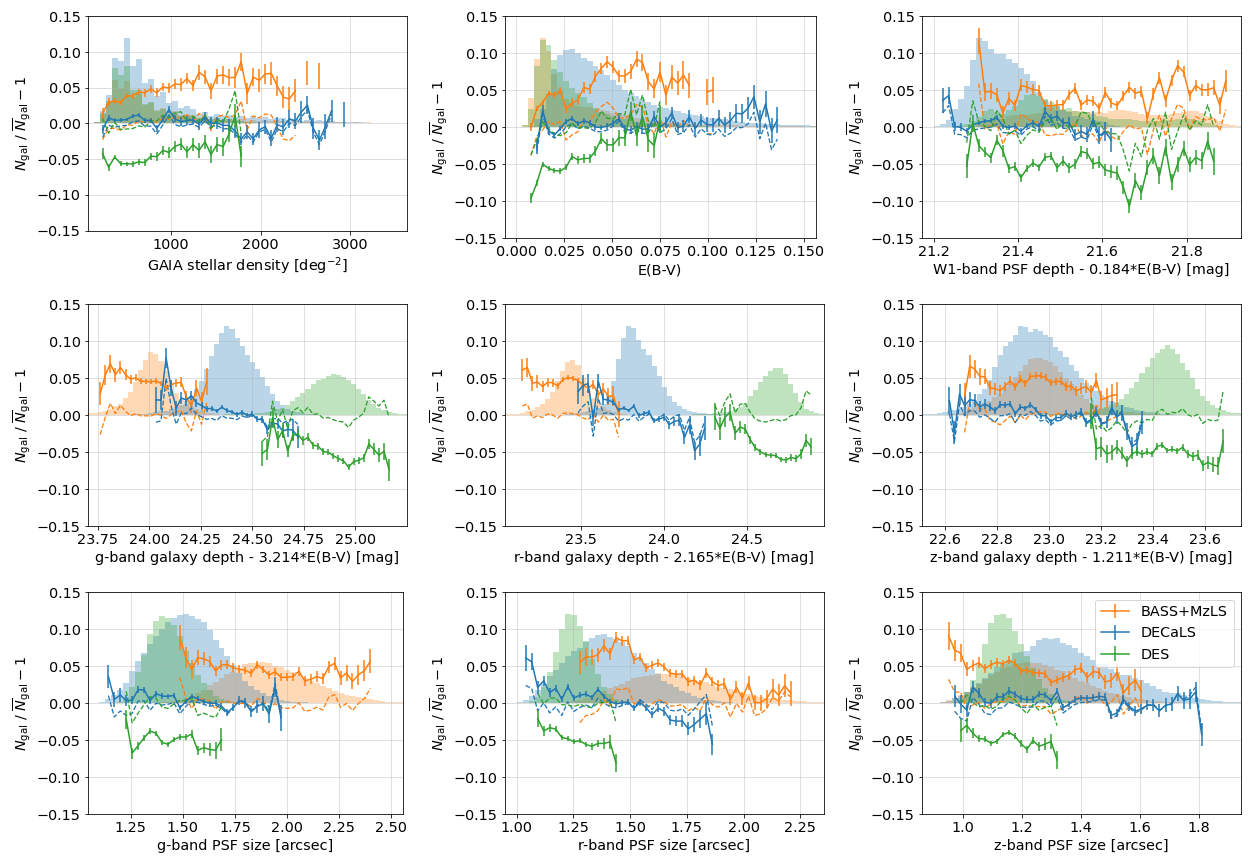}}
    \caption{As figure \ref{fig:main_lrg_density_trends} but for the same (4th) redshift bin of the Extended sample.}
    \label{fig:extended_lrg_density_trends}
\end{figure}

Figures \ref{fig:main_lrg_density_trends} and \ref{fig:extended_lrg_density_trends} show the density trends with imaging/foreground systematics for the 4th redshift bin of the Main and Extended samples, respectively. The 4th redshift bin is shown here because it is the faintest subsample and has some of the largest density trends. To mitigate the imaging systematics, we provide weights based on linear regression fits of the systematics properties as described in section 3.1 of \cite{desi_lrg_paper}. 
The imaging weights are based on the assumption that the variations in the galaxy number densities due to imaging variations are effectively the result of random downsampling of the galaxy sample, and the multiplicative weights correct for the downsampling factor. The systematics properties used in the linear regression are seeing and depth in three optical bands and $E(B-V)$. The ``galaxy depth'' values in the LS DR9 catalogs do not account for Galactic extinction which effectively reduces the imaging depth for extragalactic sources, and to account for this effect we add an $E(B-V)$ term to the depth value. While the WISE $W1$ magnitude is used in the LRG selection, we do not include the $W1$ depth as an input for the linear regression because 1) we do not see a significant density trend with $W1$ depth and 2) there is a discrepancy in the value of the $W1$ depth between the tractor catalog and the randoms that prevent us from using the randoms to assign weights.

\begin{figure}
    \hspace*{-1.4cm}
    \resizebox{0.565\columnwidth}{!}{\includegraphics{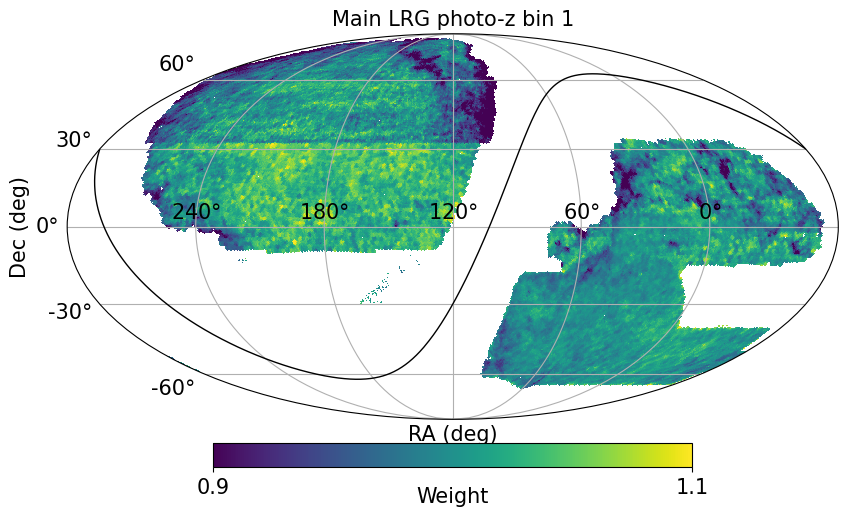}}
    \resizebox{0.565\columnwidth}{!}{\includegraphics{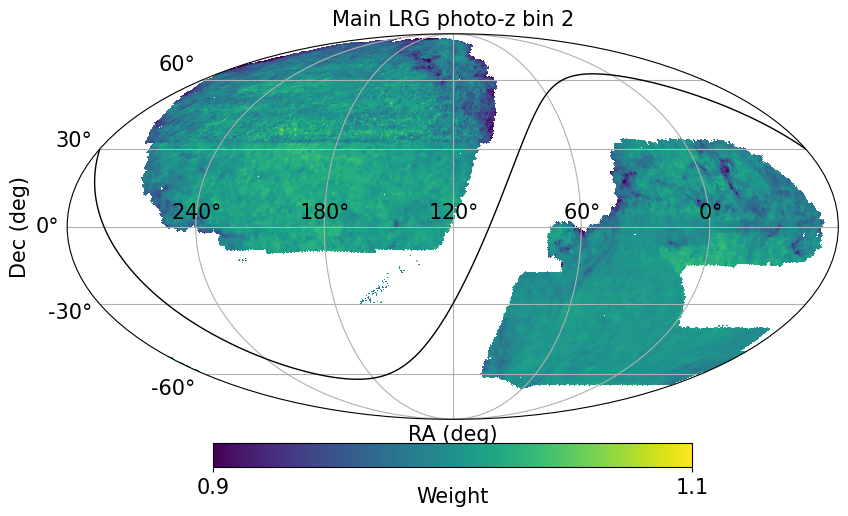}}
    \hspace*{-1.4cm}
    \resizebox{0.565\columnwidth}{!}{\includegraphics{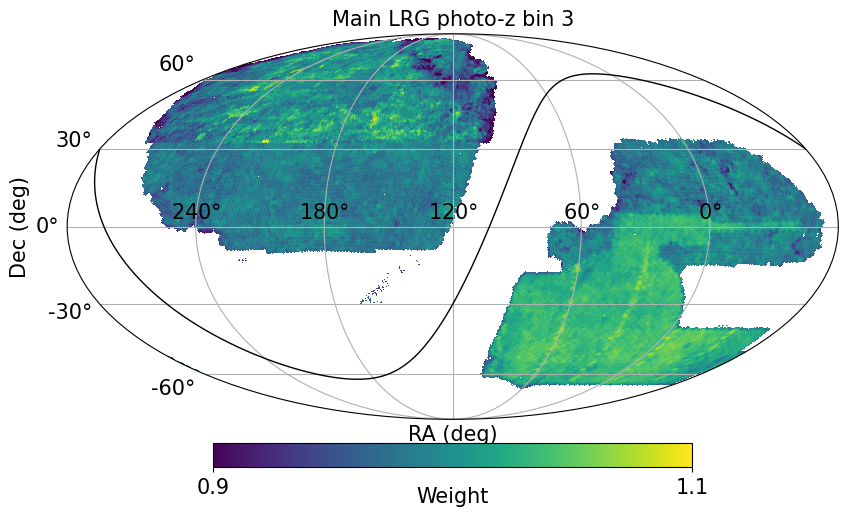}}
    \resizebox{0.565\columnwidth}{!}{\includegraphics{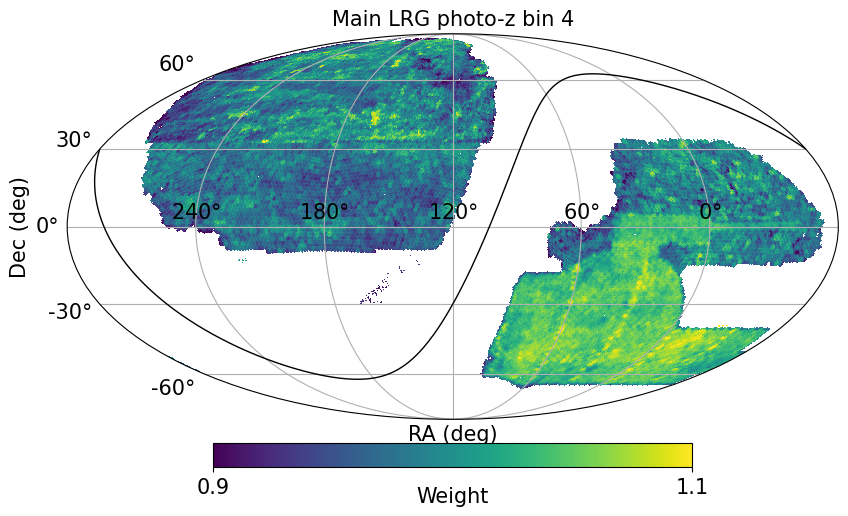}}
    \caption{Maps of the imaging systematics weights for the Main LRG subsamples that include $E(B-V)$ in the linear regression variables, shown in HEALPix resolution of NSIDE=128.}
    \label{fig:main_weight_maps}
\end{figure}

The density trends after applying the linear regression weights are shown as the dashed lines in figures \ref{fig:main_lrg_density_trends} and \ref{fig:extended_lrg_density_trends}. Figure \ref{fig:main_weight_maps} shows the weights maps (i.e., the average weight of objects within each pixel) for the Main LRGs. Some prominent features in the weight maps include variations in Galactic extinction $E(B-V)$, e.g., in the North Polar Spur (at $\mathrm{RA}\sim 150^{\circ}, \mathrm{Dec}\sim 70^{\circ}$), and depth and seeing variations, e.g., the exposure tiling patterns and the overall offsets in the deeper DES region and across the North/South boundary.

\begin{figure}
    \hspace*{-1.4cm}
    \resizebox{0.565\columnwidth}{!}{\includegraphics{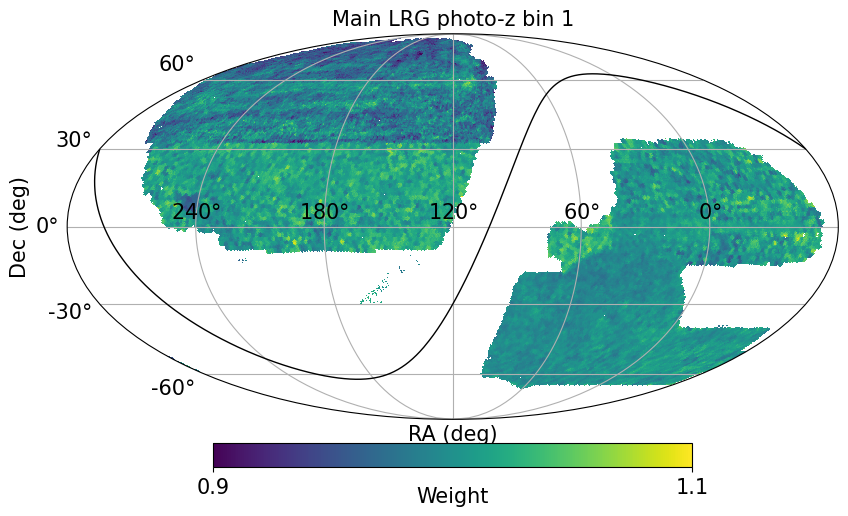}}
    \resizebox{0.565\columnwidth}{!}{\includegraphics{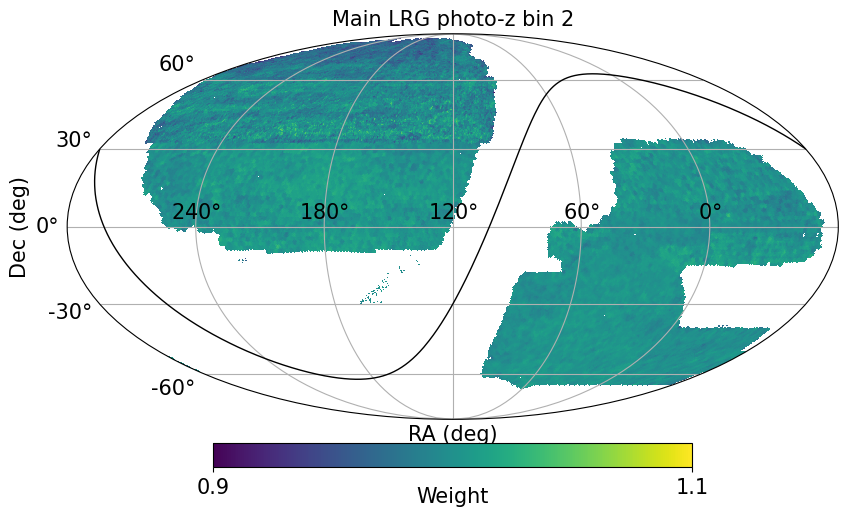}}
    \hspace*{-1.4cm}
    \resizebox{0.565\columnwidth}{!}{\includegraphics{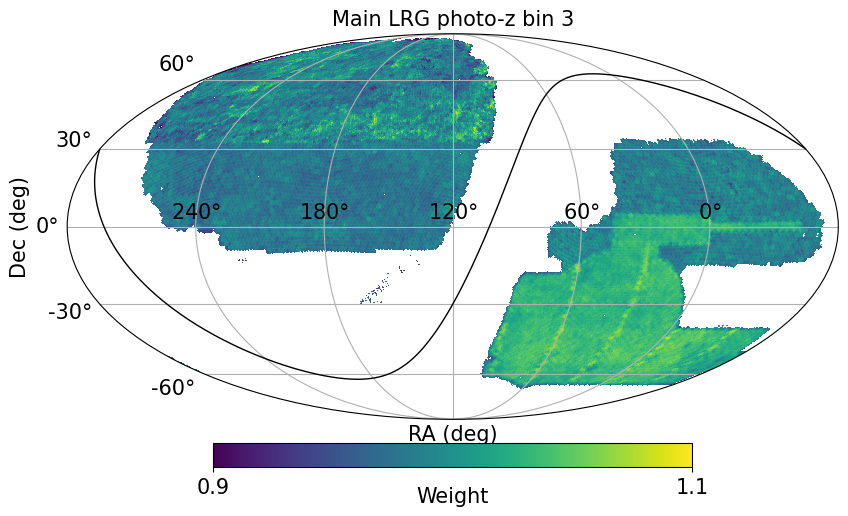}}
    \resizebox{0.565\columnwidth}{!}{\includegraphics{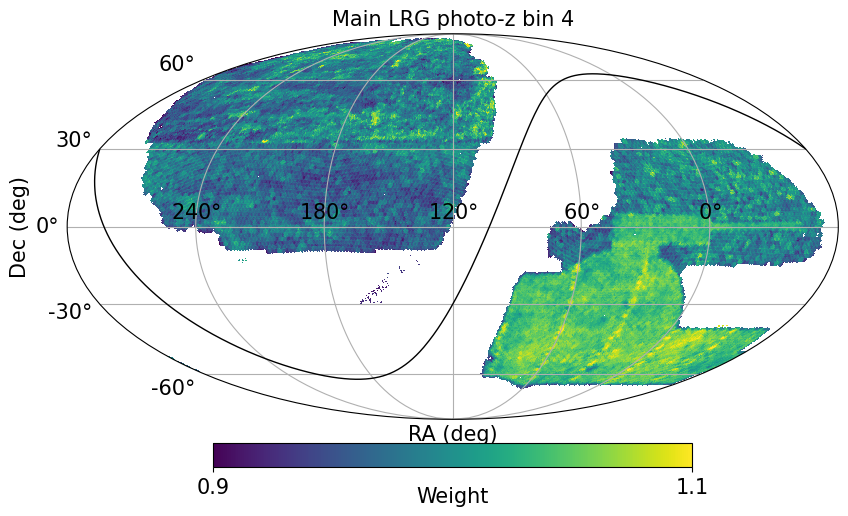}}
    \caption{As figure \ref{fig:main_weight_maps} but without $E(B-V)$ as input for the imaging systematics weights.}
    \label{fig:main_weight_maps_no_ebv}
\end{figure}

It is known that the $E(B-V)$ map from \cite{schlegel_maps_1998} is correlated with large-scale structures due to contamination from the cosmic infrared background (e.g., see \cite{chiang_extragalactic_2019}), and the inclusion of $E(B-V)$ in the linear regression variables might remove actual LSS clustering signal. Therefore we also provide a separate set of weights that do not use $E(B-V)$ in the linear regression fits. We show the weight maps without $E(B-V)$ in figure \ref{fig:main_weight_maps_no_ebv}. For the angular power spectrum and correlation function calculations in section \ref{sec:clustering_and_forecasts}, we use the weights that use $E(B-V)$. For the Main LRGs, the changes in $C_\ell$ from using the $E(B-V)$-free weights are mainly at large angular scales, and they are relatively small (only reaching 1\% for the lowest-redshift subsample, and only in the lowest $\ell$ bin). The lowest-redshift subsample is most sensitive to $E(B-V)$ because it is most affected by the $g-W1$ and $r-W1$ cuts, which are sensitive to $E(B-V)$, in the Main LRG selection that removes $z<0.4$ galaxies.

\begin{figure}
    \hspace*{-1.4cm}
    \resizebox{0.565\columnwidth}{!}{\includegraphics{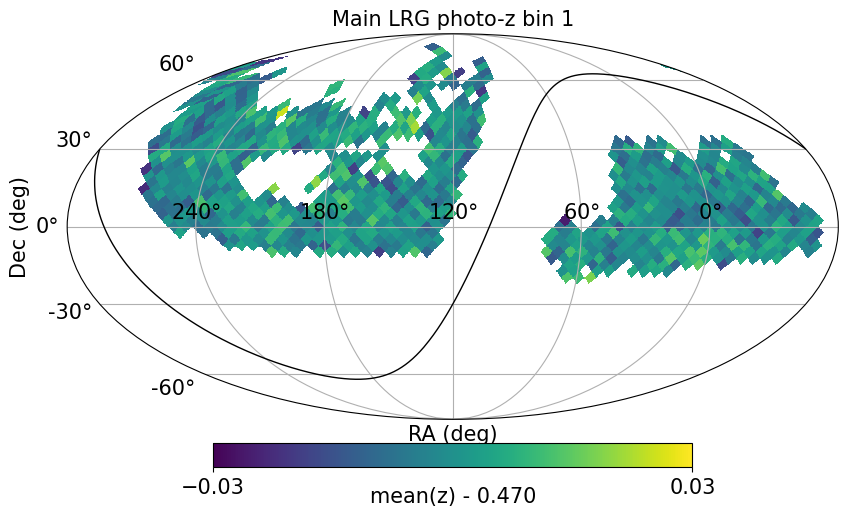}}
    \resizebox{0.565\columnwidth}{!}{\includegraphics{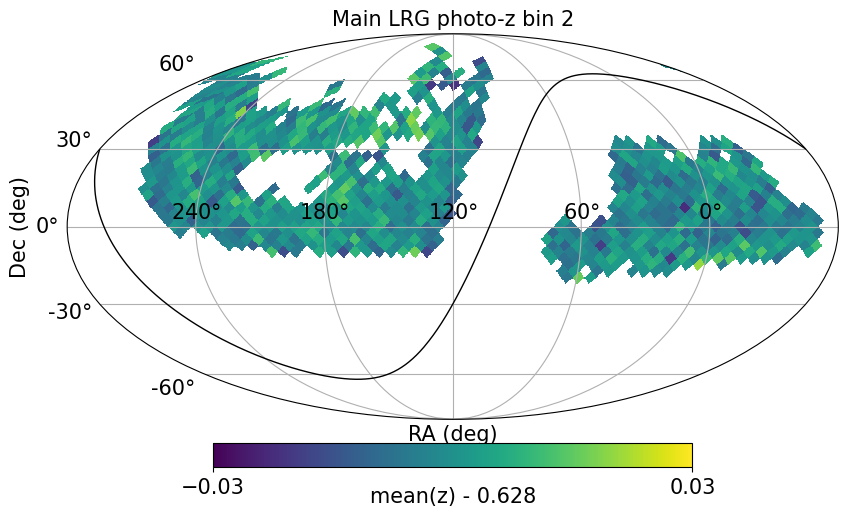}}
    \hspace*{-1.4cm}
    \resizebox{0.565\columnwidth}{!}{\includegraphics{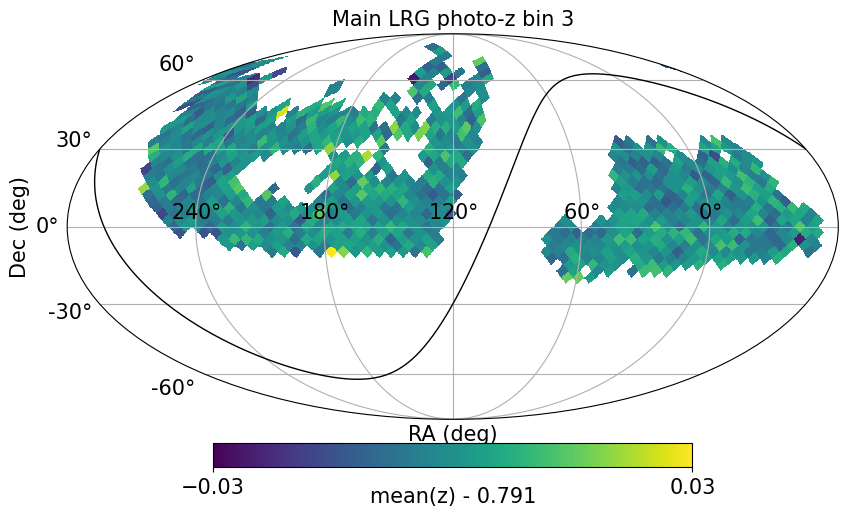}}
    \resizebox{0.565\columnwidth}{!}{\includegraphics{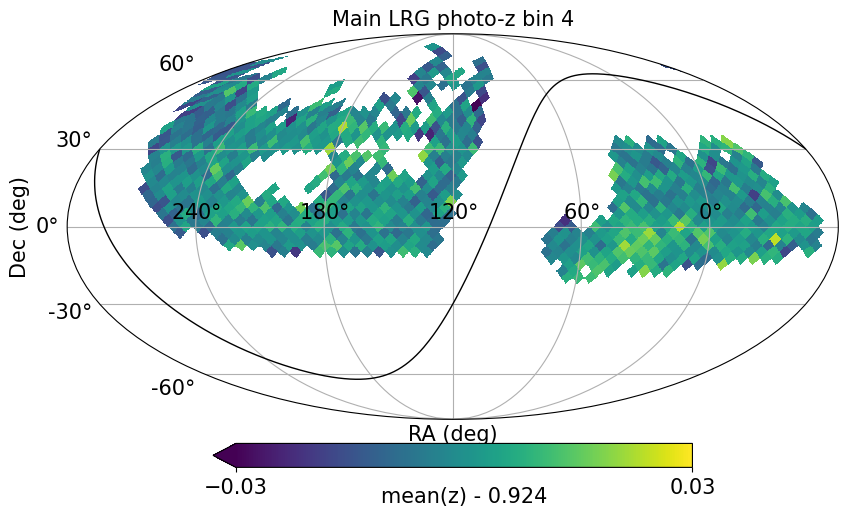}}
    \caption{Maps of the mean spectroscopic redshift for the Main LRG subsamples with HEALPix NSIDE=16, based on redshifts from the DESI Y1 data.
    }
    \label{fig:specz_mean_maps}
\end{figure}

\begin{figure}
    \hspace*{-1.4cm}
    \resizebox{0.565\columnwidth}{!}{\includegraphics{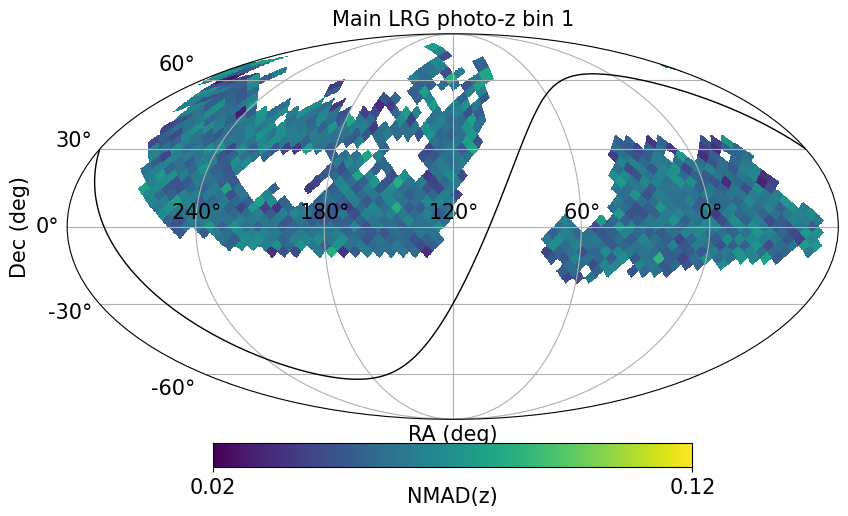}}
    \resizebox{0.565\columnwidth}{!}{\includegraphics{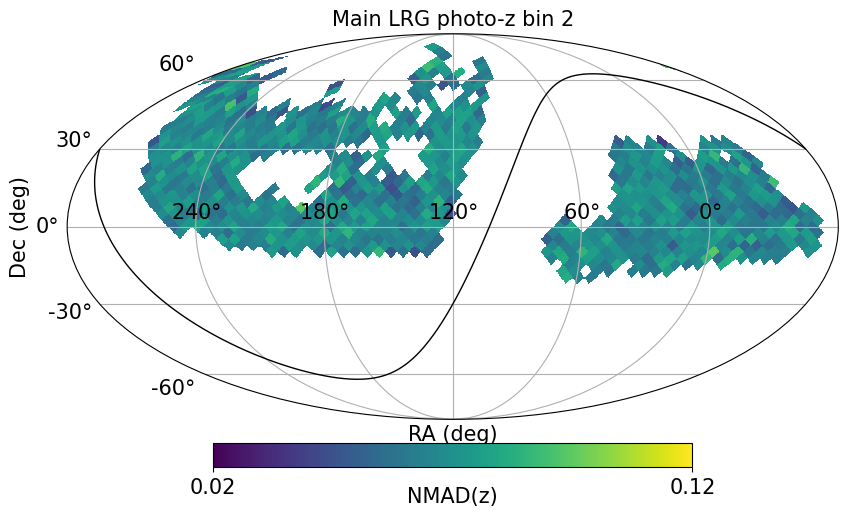}}
    \hspace*{-1.4cm}
    \resizebox{0.565\columnwidth}{!}{\includegraphics{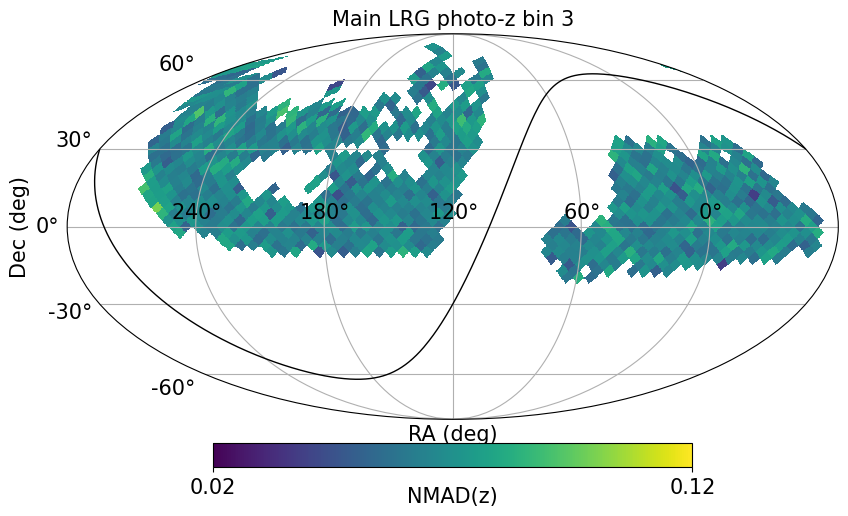}}
    \resizebox{0.565\columnwidth}{!}{\includegraphics{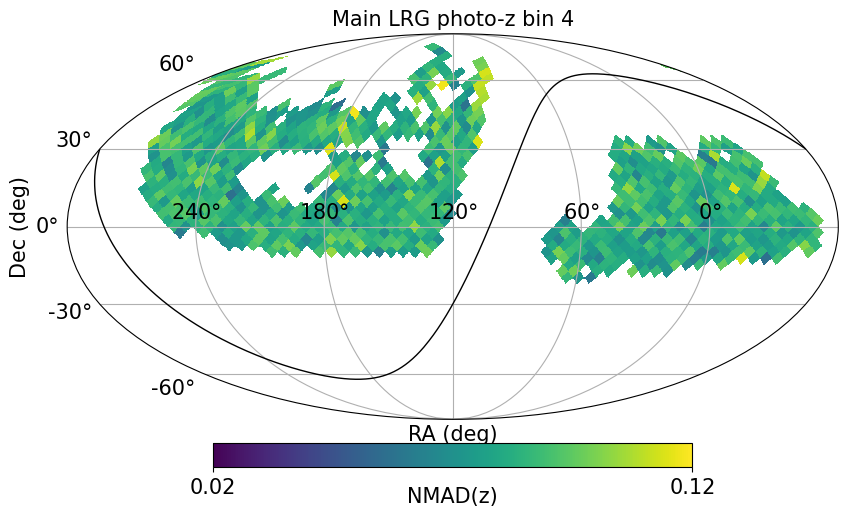}}
    \caption{Similar to figure \ref{fig:specz_mean_maps} but showing the width of the redshift bin, i.e., the normalized median absolute deviation of the spectroscopic redshifts in each HEALPix pixel (with NSIDE=16) for the Main LRG subsamples.
    }
    \label{fig:specz_nmad_maps}
\end{figure}

\begin{figure}
    \centering
    \resizebox{0.6\columnwidth}{!}{\includegraphics{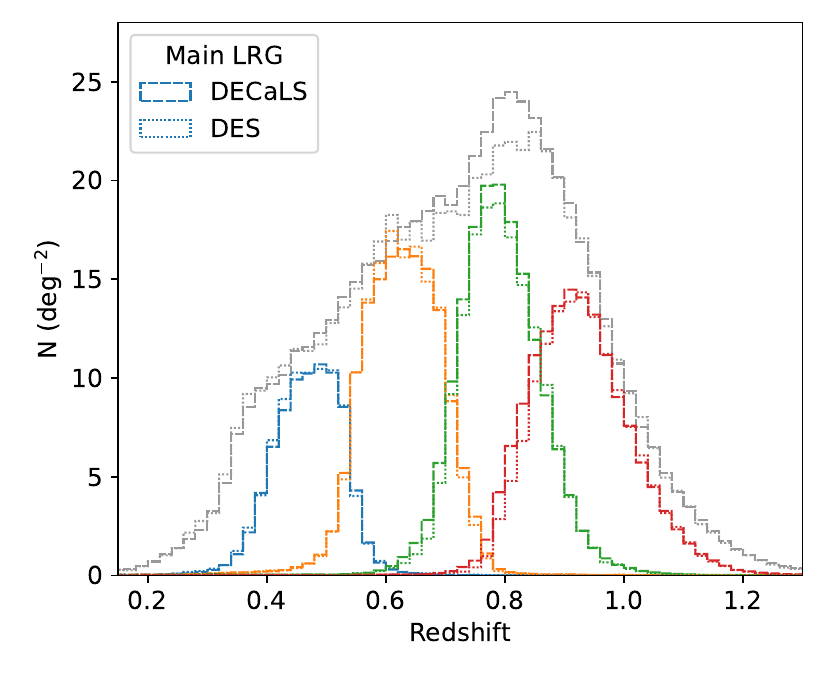}}
    \caption{As the left panel of figure \ref{fig:dndz} but showing the Main LRG redshift distributions for the DECaLS and DES regions separately. They are based on 1.67 million and 0.204 million DESI redshifts in the DECaLS and DES regions, respectively.}
    \label{fig:dndz_decals_vs_des}
\end{figure}

Besides the variation in surface density, imaging systematics could also cause spatial variations in the redshift distribution. The DESI Y1 data allows us to directly assess such variations in the Main LRG subsamples. We show maps of the mean spectroscopic redshift in each HEALPix pixel (with NSIDE=16) for each subsamples using Y1 data in figure \ref{fig:specz_mean_maps}, and maps of the ``width'' (measured by the normalized median absolute deviation) of the redshift bins in figure \ref{fig:specz_nmad_maps}. 
Both sets of maps are almost entirely dominated by noise. To interpret the results, we compare the DESI LRG subsamples with 25 sets of Abacus mock LRG catalogs \cite{first_gen_mocks_paper} (also see \cite{moon_first_2023}) based on the AbacusSummit simulation \cite{maksimova_abacussummit_2021,garrison_textttabacus_2021} with the same footprint as the Y1 data. The mock LRG subsamples are created by randomly sub-sampling the mock catalogs to reproduce the redshift distributions of the DESI LRG subsamples. We find that the DESI LRG subsamples have $1\%-14\%$ larger RMS in the per-pixel mean redshift than the mock realization with the largest RMS. For the RMS scatter in the per-pixel redshift bin width, they agree within $4\%$. The larger per-pixel mean redshift scatter in the LRG subsamples may be partly due to a mismatch between the mocks and the LRGs (e.g., it is known that the mocks have $\sim5\%$ lower bias than the LRG sample; see \cite{moon_first_2023}), or it may be an indication of the effects of imaging systematics on the redshift distributions (e.g., the redshift distribution for some of the redshift bins are visibly different in each imaging regions in figures \ref{fig:dndz} and \ref{fig:dndz_decals_vs_des}).

As is apparent in figures \ref{fig:main_lrg_density_trends}-\ref{fig:main_weight_maps_no_ebv}, the three imaging regions –– BASS+MzLS in the North and DECaLS and DES in the South –– have very different depth and seeing, and these differences have a noticeable impact on the LRG densities. 
The Northern imaging, which is based on imaging from the Bok and Mayall telescopes, has similar data quality compared to DECaLS in the $z$ band, but the $g$ and $r$ bands are 0.4 mag shallower and have significantly worse seeing. The LRG selection cuts and the photo-$z$ cuts in the North are adjusted to account for the instrument and depth differences, but it is not possible to completely homogenize the galaxy samples across the North/South boundary. The DES region is deeper than DECaLS, on average by 0.5 mag in $g$ and $z$ bands and 0.85 mag in $r$ band. The deeper DES imaging results in slightly different sample properties (e.g., surface density), especially for the two highest-redshift bins in which the LRGs are fainter and more sensitive to depth variations. 
Table \ref{tab:main_bins_three_regions} lists the average surface density, mean redshift and scatter separately in each of the three regions, and figure \ref{fig:dndz_decals_vs_des} shows the redshift distributions of the Main LRG samples in the DECaLS and DES regions separately.

We note that due to the differences in depth and seeing, the physical properties (such as mass distribution, bias etc) of the samples in the different regions are slightly different: we believe this is because shallower regions will have more scatter of fainter/lower mass galaxies into the selection boundary. While variations in number density can be corrected with the systematic weights as described in this section, changes in physical properties cannot. Depending on the application and on the required accuracy, one may want to treat the galaxies in the DES region as a different sample and fit bias parameters independently. For some applications, measuring the correct mean bias is sufficient and the split is not necessary. A full study of the differences is beyond the scope of this paper, but at fixed cosmology and using the same scale cuts as \cite{White:2021yvw}, the best-fit linear bias parameters in the DES region compared to the rest can differ by up to\footnote{We find strong evidence for a difference in clustering power for bin 4, and mild evidence for bins 2 and 3, while bin 1 is consistent with having the same clustering within the measurement error.} $\sim$ 0.7, 2, 3, 5\% in the four photometric bins. At the same time, table \ref{tab:main_bins_three_regions} and figure \ref{fig:dndz_decals_vs_des} show that there some differences in $\bar{z}$, which are at the 0.001 level for the first two bins, but are $\sim 0.003$ and $\sim 0.007$ for bins 3 and 4 respectively. Depending on the analysis being performed, this difference can have non-negligible impact on the clustering (especially when considering the full redshift distribution on the two footprints, rather than just $\bar{z}$) and may need to be taken into account. In summary, the highest redshift bins show statistically significant differences in the value of the bias and $\bar{z}$ and therefore the sample in the DES region and in other regions should be considered as different samples with different redshift distributions and independent bias parameters and shot noise for many applications. Clustering differences between BASS+MzLS in the North and DECaLS in the South appear negligible\footnote{however we have not performed full cosmological fits. Nonetheless, they are much smaller than the difference between the DES footprint and the remaining area.}, so depending on the application and the accuracy required, they may be treated as a single sample.

\begin{table}[]
    \centering
    \begin{tabular}{c|ccc|ccc|cccc}
         Bin & & $\bar{n}_\theta$ & & & $\bar{z}$ & & & $\Delta z$ & \\
         \hline
           & North & DECaLS & DES & North & DECaLS & DES & North & DECaLS & DES & \\
         1 &  83.3 &  80.9 &  82.4 & 0.470 & 0.470 & 0.469 & 0.064 & 0.063 & 0.064 \\
         2 & 149.3 & 147.5 & 147.9 & 0.630 & 0.627 & 0.626 & 0.075 & 0.073 & 0.072 \\
         3 & 163.7 & 164.7 & 157.4 & 0.791 & 0.791 & 0.794 & 0.079 & 0.078 & 0.077 \\
         4 & 150.2 & 151.0 & 142.3 & 0.922 & 0.925 & 0.932 & 0.099 & 0.094 & 0.092
    \end{tabular}
    \caption{Main LRG statistics similar to table \ref{tab:main_bins} but separately listing the values for the three imaging regions: North (BASS+MzLS), DECaLS and DES.}
    \label{tab:main_bins_three_regions}
\end{table}

\section{Number count slope and magnification bias}
\label{sec:slope}

In addition to the ``intrinsic'' clustering of galaxies that arises from their tracing of the cosmic web, the observed galaxy number count is also modulated by a lensing magnification term, which correlates the observed number count with matter along the line of sight. Knowing the size of the magnification bias, parameterized by the ``number count slope'' defined below, is necessary for interpreting both the galaxy auto-correlation and its cross-correlation with other tracers (see for example \cite{Krolewski:2019yrv} for its importance in recent analyses). For the purpose of estimating magnification bias, we measure the response of the number density of galaxies in each redshift bin to lensing magnification. This response is quantified as the in logarithmic change in the galaxy number count for a small change in magnitude (in all bands) due to magnification,
\begin{equation}
    s \equiv d\log_{10}N/dm . \label{eq:number_count_slope}
\end{equation}
For a magnitude-limited sample, $s$ is simply the ``slope'' of the (logarithmic) number count as a function of the limiting magnitude. While our selection is not a magnitude-limited selection, we nonetheless adopt the term ``number count slope'' for $s$. Note that the magnification slope $s$ only applies to the sample in question as a whole, and the value is expected to change if we select a subsample, for example by color. We determine the value of $s$ using finite differences, by shifting all magnitudes by a small amount, $\delta m = \pm0.01$, and re-applying the selection. However, there are two subtleties specific to our LRG samples that prevent us from simply using Equation \ref{eq:number_count_slope} for computing $s$.

First, the LRG selections include fiber-magnitude cuts such as Eq. \ref{eq:mag-limit}, which is effectively a cut on surface brightness. If a galaxy is magnified, the observed (total) magnitude changes by $\delta m$, but the fiber-magnitude does not change by the same amount. This is because, for an extended object (i.e., a resolved galaxy), the fiber-flux fraction (the fraction of flux within the fiber diameter) decreases as the size of the galaxy increases. Therefore, to properly model the response of the number density of galaxies to lensing magnification, one needs to know how the fiber-flux (or, equivalently, the fiber-flux fraction) of each galaxy changes when it is magnified or demagnified. We measure the fiber-flux fraction as a function of the shape parameters for each morphology type, compute the fractional (fiber-)flux change when the half-light radius is changed by a small amount, and use it to compute $\delta m_\mathrm{fiber}$ for a given magnification-induced $\delta m$. See appendix \ref{app:fibermagnitude} for details of how $\delta m_\mathrm{fiber}$ is computed.

The other subtlety is that the redshift binning is done using photo-$z$'s. Galaxy colors do not change with magnification, but the random forest-based photo-$z$'s use magnitudes and galaxy sizes as features, which do respond to magnification. To account for the change in photo-$z$'s, we recompute the photo-$z$'s with the shifted magnitudes and sizes. To reduce random noise from the random forest realizations, we use the same random seed for computing the magnified and unmagnified photo-$z$'s.

Now that we have the $\delta m_\mathrm{fiber}$ value and the ``shifted'' photo-$z$'s for each galaxy for magnitude shifts of $\delta m = \pm0.01$, we reapply the LRG selection and the photo-$z$ binning. The resulting value of the number count slope $s$ from finite difference for each subsample is listed in tables \ref{tab:main_magnification} and \ref{tab:extended_magnification} for the Main and Extended samples, respectively. We list the values for the North+South combined sample, and for North and South separately. (We do not list the values for DECaLS and DES separately because their differences are small and well within the statistical uncertainties.) To demonstrate the effects of the fiber-flux cut and photo-$z$ binning, we also list $s$ values if we let the fiber-magnitude change by the same amount as the total magnitude, or if we use the ``unshifted'' photo-$z$'s for the photo-$z$ re-binning. The slope would be overestimated by ${\sim}\,13\%$ for the unbinned Main LRG sample if the fiber-flux effects are not accounted for, or ${\sim}\,30\%$ for the highest redshift bin (which is also the faintest bin and most subject to the fiber-flux cut). Shifts in the photo-$z$'s also have a similarly significant effect on $s$. The actual impact from the magnification term is even larger than these fractional changes in $s$ since the magnitude kernel scales with $s-0.4$ (e.g., see equation 3.11 of \cite{krolewski_cosmological_2021}).

\begin{table}
\centering
\begin{tabular}{c|ccc|c|c}
Bin & North & South & \textbf{Combined} & $dm_\mathrm{fiber}=dm$ & Same photo-$z$ \\
\hline
all   & $1.021 \pm 0.012$ & $1.001 \pm 0.008$ & $\boldsymbol{1.008 \pm 0.007}$ & $1.136 \pm 0.007$ & $1.008 \pm 0.007$ \\
1 & $1.002 \pm 0.033$ & $0.958 \pm 0.022$ & $\boldsymbol{0.972 \pm 0.018}$ & $0.979 \pm 0.018$ & $0.997 \pm 0.018$ \\
2 & $1.055 \pm 0.025$ & $1.039 \pm 0.017$ & $\boldsymbol{1.044 \pm 0.014}$ & $1.073 \pm 0.014$ & $1.007 \pm 0.014$ \\
3 & $0.964 \pm 0.024$ & $0.978 \pm 0.016$ & $\boldsymbol{0.974 \pm 0.013}$ & $1.089 \pm 0.013$ & $0.862 \pm 0.013$ \\
4 & $1.015 \pm 0.025$ & $0.976 \pm 0.016$ & $\boldsymbol{0.988 \pm 0.014}$ & $1.284 \pm 0.014$ & $1.148 \pm 0.014$ \\
\end{tabular}
\caption{\label{tab:main_magnification} The magnification ``number count slope'' value $s$ for the Main LRG sample, and its Poisson error from the finite difference calculation with $\delta m = \pm0.01$. The ``all'' row is for the full DESI LRG sample (including objects that are not in any of the redshift bins). The ``North'', ``South'' and ``Combined'' columns are our best estimates that correct for the impact of fiber-flux and photo-$z$'s. To show how large these effects are, the last two columns list the incorrect estimates (for north and south combined) that do not correct for either of these effects; these values are for illustration only and should not be used. We recommend using the values in the ``Combined'' column (in boldface) for the North+South combined sample.}
\end{table}

\begin{table}
\centering
\begin{tabular}{c|ccc|c|c}
& North & South & \textbf{Combined} & $dm_\mathrm{fiber}=dm$ & Same photo-$z$ \\
\hline
all   & $0.752 \pm 0.007$ & $0.748 \pm 0.005$ & $\boldsymbol{0.750 \pm 0.004}$ & $0.846 \pm 0.004$ & $0.750 \pm 0.004$ \\
1 & $0.805 \pm 0.022$ & $0.776 \pm 0.015$ & $\boldsymbol{0.785 \pm 0.012}$ & $0.787 \pm 0.012$ & $0.839 \pm 0.012$ \\
2 & $0.862 \pm 0.017$ & $0.832 \pm 0.011$ & $\boldsymbol{0.841 \pm 0.009}$ & $0.863 \pm 0.009$ & $0.790 \pm 0.009$ \\
3 & $0.655 \pm 0.014$ & $0.677 \pm 0.010$ & $\boldsymbol{0.670 \pm 0.008}$ & $0.750 \pm 0.008$ & $0.556 \pm 0.008$ \\
4 & $0.681 \pm 0.014$ & $0.683 \pm 0.010$ & $\boldsymbol{0.682 \pm 0.008}$ & $0.868 \pm 0.008$ & $0.812 \pm 0.008$ \\
\end{tabular}
\caption{\label{tab:extended_magnification} As table \ref{tab:main_magnification}, but for the Extended LRG sample.}
\end{table}

\section{Angular clustering and sample characterization}
\label{sec:clustering_and_forecasts}

\begin{figure}
    \centering
    \resizebox{0.9\columnwidth}{!}{\includegraphics{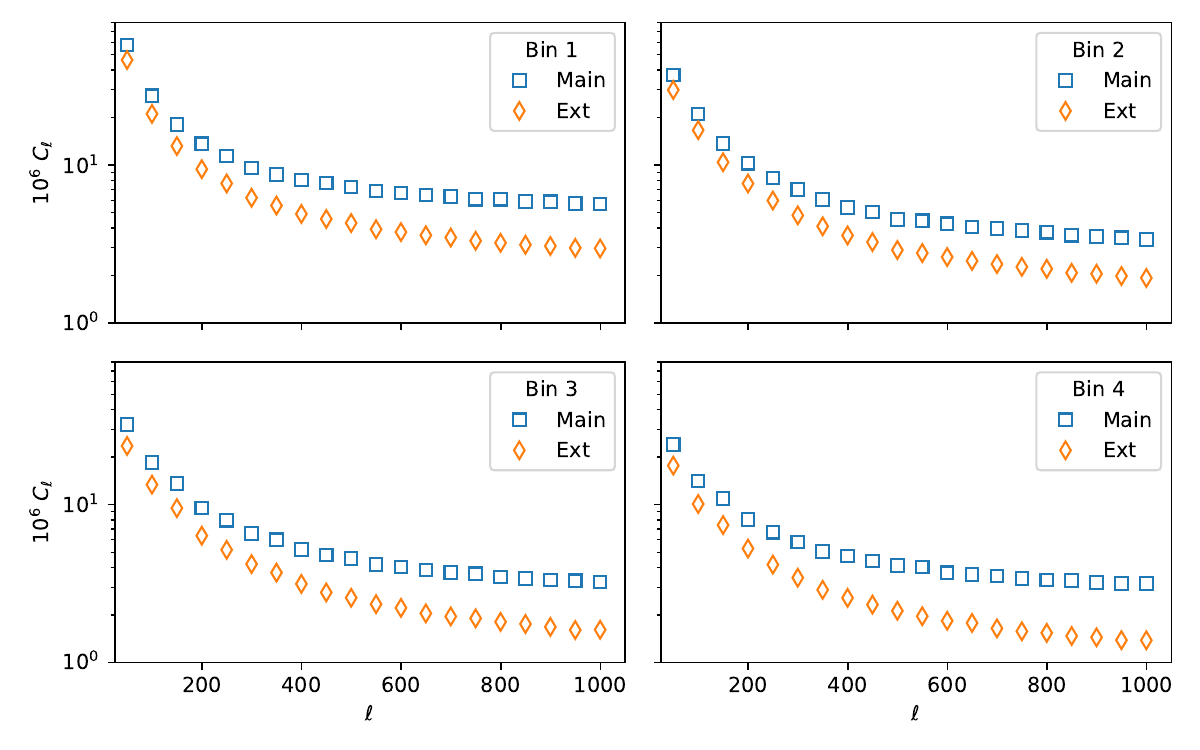}}
    \caption{Angular power spectra of the Main LRG subsamples and the Extended LRG subsamples.  The large-scale clustering of the Extended samples is only slightly lower than that of the Main sample (due to the lower bias, see text), but the small-scale power is much reduced due to the large reduction in shot noise.}
    \label{fig:c_ell}
\end{figure}

\begin{figure}
    \resizebox{0.9\columnwidth}{!}{\includegraphics{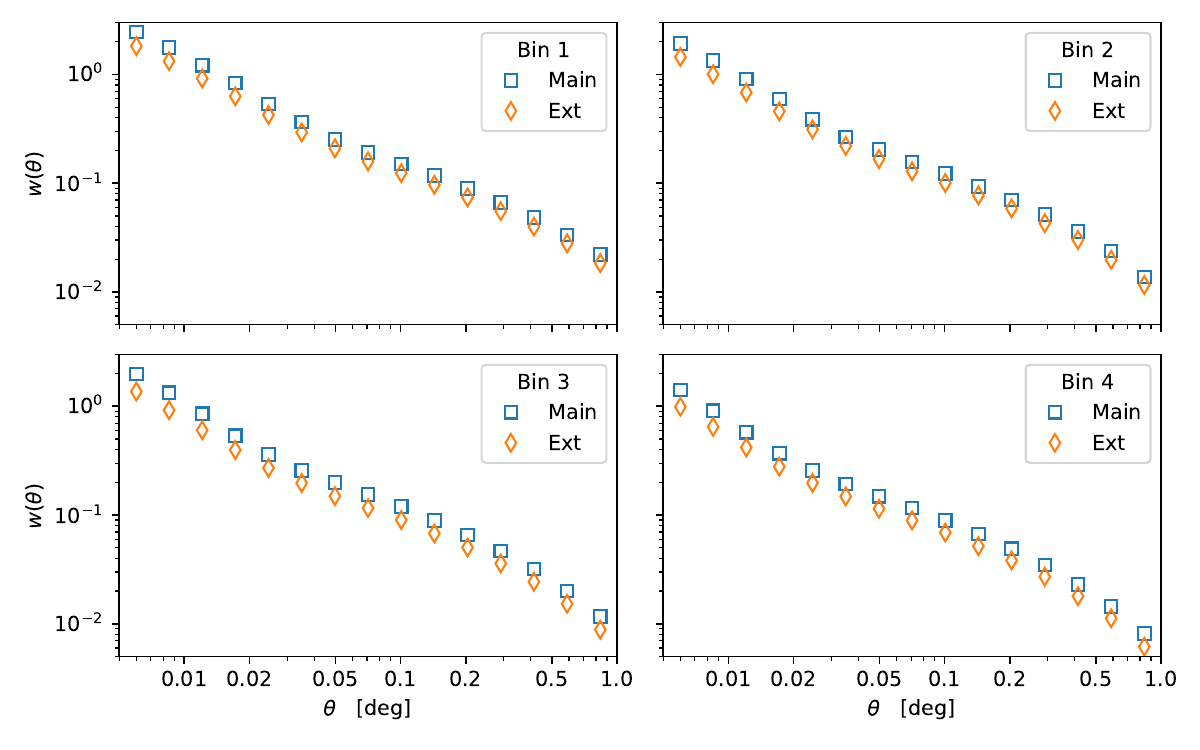}}
    \caption{Angular correlation functions of the Main LRG subsamples and the Extended LRG subsamples.  For all samples the small-scale clustering of the Main sample exceeds that of the Extended sample, though the large-scale clustering of the Extended samples is only slightly lower than that of the Main sample (due to the lower bias, see text).}
    \label{fig:wtheta}
\end{figure}

The angular power spectra and angular correlation functions for our samples are shown in figures \ref{fig:c_ell} and \ref{fig:wtheta}.  In all cases we detect the clustering at very high significance.  In the angular power spectra (figure \ref{fig:c_ell}) we see that the Main sample has slightly higher large-scale bias than the Extended sample (the low $\ell$ blue points are slightly above the red ones) and significantly higher shot-noise (the blue points at high $\ell$ are well above the red points).  This is expected as the Extended sample tends to be less luminous overall and has a higher number density.  Assuming a Planck cosmology \cite{PlanckI,PlanckVI} the large-scale bias of the Main sample, estimated from fitting the galaxy auto-spectra, ranges from 1.8-2.3 with increasing redshift.  The bias of the Extended sample is approximately 10\% lower.  The angular correlation functions (figure \ref{fig:wtheta}) tell a similar story, though the difference in shot noise is not evident in $w(\theta)$.

While we have not attempted to fit a halo occupancy distribution (HOD) to these specific samples in this paper, for the Main LRGs as a whole, fits to the small-scale correlation function \cite{LRG_HOD} find a satellite fraction of $11 \pm 1 \%$ and a mean halo mass of $\log_{10}M_h = 13.40 \pm 0.02$ at $0.4 < z < 0.6$ (roughly applicable to the lowest redshift bin) and a satellite fraction of $14 \pm 1 \%$ and $\log_{10}M_h = 13.24 \pm 0.02$ at $0.6 < z < 0.8$ (roughly applicable to the higher redshift bins). 

The combination of high and relatively constant number density with redshift in well-defined, spectroscopically calibrated redshift bins (figure \ref{fig:comoving_density}) and high bias (figure \ref{fig:c_ell}) makes these samples ideally suited for cross-correlation science.  An earlier version of the Main LRG tomographic sample \cite{White:2021yvw} was used in cross-correlation with CMB lensing measured by Planck \cite{PlanckLens18} to make a high-SNR measurement of $\sigma_8$ as a function of redshift (tomography) in the four redshift bins described here. Combining all redshift bins together, \cite{White:2021yvw} found a value of $S_8 = \sigma_8 \sqrt{\Omega_m / 0.3} = 0.73 \pm 0.03$, or a $\approx 4 \%$ measurement of the matter fluctuations at low redshift.

Moving forward, the rapid increase in CMB lensing map noise will allow for a considerable improvement in constraints compared to current analyses using Planck CMB lensing. Assuming 13,000 sq deg of overlap between the CMB maps from Chile-based CMB experiments and the LRG, we expect the cross-correlation constraints of \cite{White:2021yvw} to improve by a factor $\approx$ 1.4, 2 and 2.5 with ACT, Simons Observatory and CMB-S4 \cite{CMB-S4} respectively\footnote{when used in combination with Planck CMB lensing maps and assuming $\ell_{\rm max} = 400$ throughout. Larger improvements are expected with improved modeling if $\ell_{\rm max}$ can be increased, and when combining with the CMB lensing auto power spectrum.}, leading to percent-level constraints on the amplitude of fluctuations at low redshift and shedding light on dark energy \cite{Yu:2021vce} and neutrino masses \cite{Yu:2018tem}, testing our understanding of gravity and weighing in on the ``low $S_8$'' discrepancy \cite{Abdalla:2022yfr}. 

Similarly, these samples will allow for much improved constraints on the gas properties of the halos hosting the LRG galaxies through the thermal and kinematic Sunyaev-Zel'dovich effects (tSZ and kSZ respectively). Using the assumptions and techniques outlined in \cite{Battaglia:2017neq}, we forecast that for the tSZ effect, which measures the integrated thermal pressure in the host halos, we will achieve a combined SNR $\approx 50$, 85 and 140 with ACT, Simons Observatory and CMB-S4 respectively (although care will need to be exercised to distinguish the tSZ signal from the dust emission associated with the halos themselves). This is thanks to the large number density of the LRG samples, which will also allow for the study of the redshift dependence of the signal, together with improvements in the depth of CMB temperature maps. 
Moreover, the gas density around the LRG host halos, measured through kSZ, can be detected with SNR $\approx 35$, 60 and 100 in cross-correlation with ACT, Simons Observatory and CMB-S4 respectively when DESI spectroscopy is available on the overlap region (here assumed to be $\approx$ 9,000 sq deg between Chile-based CMB experiments and DESI spectroscopy).
The combination of kSZ, tSZ and lensing will reveal the full thermodynamic properties of the halo to unprecedented precision and help us directly calibrate the baryon effects on the matter distribution, one of the main systematic uncertainties in weak lensing.

Furthermore, the spectroscopically calibrated samples described here will provide an interesting alternative to the photometric clustering samples used in many combined galaxy weak lensing and clustering analyses, e.g.\ \cite{DES:2021wwk}. The samples defined here have full overlap with the DES footprint, and cross-correlations with the third and fourth redshift bins of the DES \texttt{Metacalibration} sample \cite{Gatti2021} yield a SNR of $\approx 50$ to $\ell=600$ \cite{des_xcorr}. The overlap with the Rubin Observatory LSST Wide-Fast-Deep (WFD) survey will be approximately 10,000 sq deg, and 14,000 sq deg with Euclid, and so we expect this sample will continue to be useful for galaxy lensing cross-correlations for the foreseeable future.

\section{Data products}
\label{sec:data_products}

The data and catalogs are publicly available at \url{https://data.desi.lbl.gov/public/papers/c3/lrg_xcorr_2023/}, where README files explain the data format.
The data products include:
\begin{itemize}
  \item Main and Extended LRG catalogs that include the tomographic bin indices, veto masks, imaging systematics weights, and photo-$z$'s.
  \item Overdensity maps and masks of the Main and Extended LRG subsamples at NSIDE=2048 resolution.
  \item Lower resolution (NSIDE=256) maps of density and imaging systematics properties described in section \ref{sec:imaging_systematics}.
  \item Coefficients for the linear regression imaging systematics weights for each tomographic bin.
  \item LRG veto mask values for the randoms (row-by-row matched to the already released randoms catalogs\footnote{\url{https://data.desi.lbl.gov/public/edr/target/catalogs/dr9/0.49.0/randoms/resolve/}}).
  \item Redshift distributions in figures \ref{fig:dndz} and \ref{fig:dndz_decals_vs_des}.
  \item Angular clustering measurements in figures \ref{fig:c_ell} and \ref{fig:wtheta}.
\end{itemize}

We also make public the new set of photo-$z$'s (which are used in this work) for all DR9 objects; they are available from the Legacy Surveys website: \url{https://www.legacysurvey.org/dr9/files/#photo-z-sweeps-9-1-photo-z-sweep-brickmin-brickmax-pz-fits}.

Data points in the figures are also available at \url{https://zenodo.org/record/8319955}.

\section{Conclusions}
\label{sec:conclusions}

We presented two photometric LRG samples in tomographic bins with well-characterized redshift distributions from spectroscopy. The first sample is identical to the DESI LRGs, and the second sample has more extended selection cuts that yield 2-3 times the DESI LRG density. The photometric redshifts used for the tomographic binning are trained using DESI redshifts in addition to redshifts from other surveys, and the photo-$z$'s for all objects in LS DR9 (which includes but are not limited to the LRG samples) are publicly available. Unlike typical photometric galaxy samples that have large uncertainties in their redshift distributions, the LRG samples have the unique advantage of having DESI spectroscopic redshifts with high completeness over large areas that make $N(z)$ uncertainties negligible, and spectroscopy also confirms the very low ($0.3\%$) stellar contamination rate. 

We examined the trends in the surface density with imaging/foreground systematics, and we find that while significant trends exist, they can mostly be removed with simple weighting based on linear regression. However we find that due to the different depths and seeing between the DES region and the remaining area, the samples in these two regions may have slightly different physical properties (such as bias, mass distribution and redshift distribution), as discussed in section \ref{sec:imaging_systematics}, especially for the highest redshift bins. Depending on the analysis being performed, the galaxies in the two footprints may need to be treated as separate samples with different bias parameters, shot noise and redshift distribution (or at least consistency of the results obtained with the different samples should be checked).

For lensing and clustering analyses it is important to account for magnification bias, and for this purpose we compute the effective ``number count slope'' that properly accounts for the effects of the surface brightness cut in the selection and the photo-$z$ binning.

The sample covers a large fraction of the extragalactic sky and thus is ideal for cross-correlation with other data sets. It overlaps with the entirety of the existing weak lensing surveys DES and HSC \cite{aihara_hyper_2018}, and has 13,000 sq deg of overlap with the ACT CMB survey, 10,000 sq deg with the planned Rubin Observatory LSST WFD survey, and 14,000 sq deg with Euclid, thus enabling transformative joint-survey science.

\acknowledgments

We would like to thank Alex Krolewski and Ashley Ross for helpful discussions during the preparation of this manuscript. R.Z. and S.F. are supported by the Director, Office of Science, Office of High Energy Physics of the U.S. Department of Energy under Contract No.\ DE-AC02-05CH11231.
M.W.~is supported by the DOE and the NSF.
N.S. is supported by the Office of Science Graduate Student Research (SCGSR) program administered by the Oak Ridge Institute for Science and Education for the DOE under contract number DE‐SC0014664. 
This research has made use of NASA's Astrophysics Data System and the arXiv preprint server.

This material is based upon work supported by the U.S. Department of Energy (DOE), Office of Science, Office of High-Energy Physics, under Contract No. DE-AC02-05CH11231, and by the National Energy Research Scientific Computing Center, a DOE Office of Science User Facility under the same contract. Additional support for DESI was provided by the U.S. National Science Foundation (NSF), Division of Astronomical Sciences under Contract No. AST-0950945 to the NSF's National Optical-Infrared Astronomy Research Laboratory; the Science and Technologies Facilities Council of the United Kingdom; the Gordon and Betty Moore Foundation; the Heising-Simons Foundation; the French Alternative Energies and Atomic Energy Commission (CEA); the National Council of Science and Technology of Mexico (CONACYT); the Ministry of Science and Innovation of Spain (MICINN), and by the DESI Member Institutions: \url{https://www.desi.lbl.gov/collaborating-institutions}.

The DESI Legacy Imaging Surveys consist of three individual and complementary projects: the Dark Energy Camera Legacy Survey (DECaLS), the Beijing-Arizona Sky Survey (BASS), and the Mayall $z$-band Legacy Survey (MzLS). DECaLS, BASS and MzLS together include data obtained, respectively, at the Blanco telescope, Cerro Tololo Inter-American Observatory, NSF's NOIRLab; the Bok telescope, Steward Observatory, University of Arizona; and the Mayall telescope, Kitt Peak National Observatory, NOIRLab. NOIRLab is operated by the Association of Universities for Research in Astronomy (AURA) under a cooperative agreement with the National Science Foundation. Pipeline processing and analyses of the data were supported by NOIRLab and the Lawrence Berkeley National Laboratory. Legacy Surveys also uses data products from the Near-Earth Object Wide-field Infrared Survey Explorer (NEOWISE), a project of the Jet Propulsion Laboratory/California Institute of Technology, funded by the National Aeronautics and Space Administration. Legacy Surveys was supported by: the Director, Office of Science, Office of High Energy Physics of the U.S. Department of Energy; the National Energy Research Scientific Computing Center, a DOE Office of Science User Facility; the U.S. National Science Foundation, Division of Astronomical Sciences; the National Astronomical Observatories of China, the Chinese Academy of Sciences and the Chinese National Natural Science Foundation. LBNL is managed by the Regents of the University of California under contract to the U.S. Department of Energy. The complete acknowledgments can be found at \url{https://www.legacysurvey.org/}.

Any opinions, findings, and conclusions or recommendations expressed in this material are those of the author(s) and do not necessarily reflect the views of the U. S. National Science Foundation, the U. S. Department of Energy, or any of the listed funding agencies.

The authors are honored to be permitted to conduct scientific research on Iolkam Du'ag (Kitt Peak), a mountain with particular significance to the Tohono O'odham Nation.

\appendix

\section{Comparison with the LRG sample in Ref. \cite{White:2021yvw}}
\label{app:old_sample}
An earlier version of the Main LRG sample was used in the companion paper \cite{White:2021yvw}. It has a few notable differences compared to the updated sample presented in this work: 
1) Ref.~\cite{White:2021yvw} used an older version of photo-$z$'s for tomographic binning.
2) Ref.~\cite{White:2021yvw} used a smaller subset of the early DESI data for estimating the redshift distributions.
3) At $\mathrm{Dec}<-29.25^{\circ}$, the updated samples are based on photometry with corrected zero point offsets.
4) When requiring 2+ exposures in the $grz$ bands, we switched from the tractor-based NOBS value to the pixel-based PIXEL\_NOBS value; this results in 0.5\% fewer objects compared to the sample in \cite{White:2021yvw}.
5) The imaging systematics weights in \cite{White:2021yvw} included WISE $W1$ depth as an input, but we have learned that there is a discrepancy in the WISE $W1$ depth value between the LRGs and the randoms which could bias the weight values; for this reason we dropped W1 depth from the inputs for the imaging weights in this work.
Moreover, the differences between the DES region and other regions were not understood in \cite{White:2021yvw}, and the entire footprint was treated as a single galaxy sample per redshift bin in the analysis.

\begin{figure}
    \centering
    \resizebox{0.8\columnwidth}{!}{\includegraphics{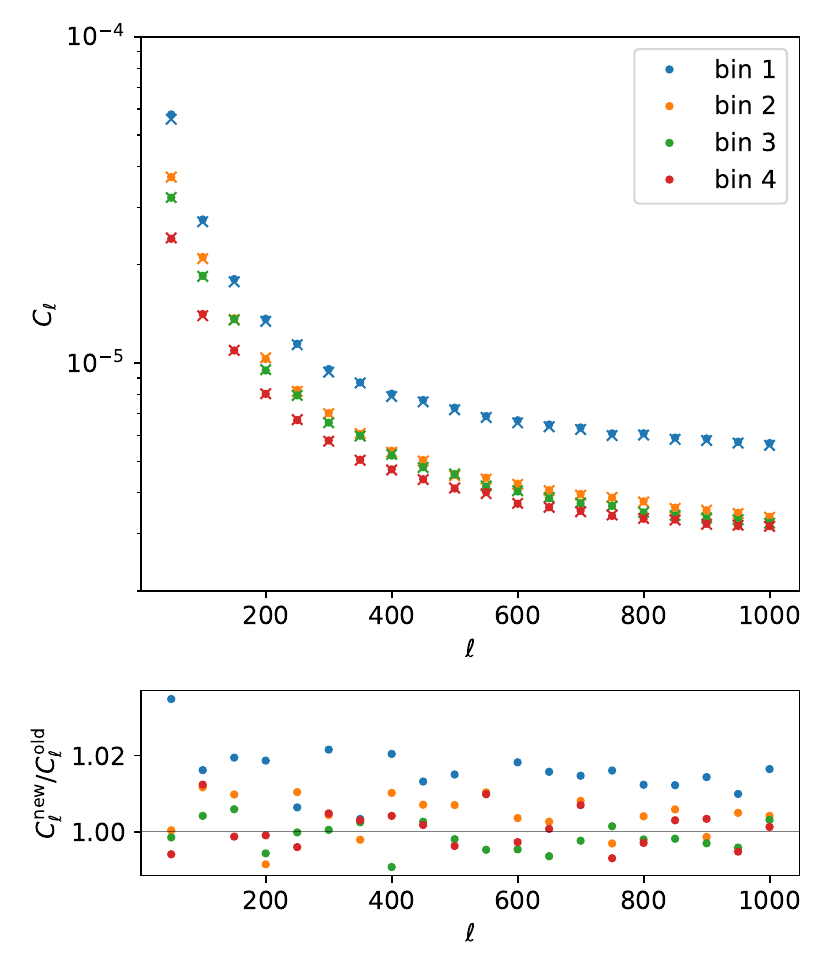}}
    \caption{Top panel: comparison of the angular power spectrum between the Main LRG subsamples in this paper (filled circles) and the older LRG subsamples used in \cite{White:2021yvw} (crosses). Bottom panel: the $C_\ell$ ratio between the two samples.
    }
    \label{fig:c_ell_new_old}
\end{figure}

Figure \ref{fig:c_ell_new_old} compares the angular power spectrum of the Main LRG subsamples with the older LRG subsamples in \cite{White:2021yvw}. The two samples mostly agree within $2\%$, although on average the new sample has higher $C_\ell$. The slightly higher $C_\ell$ is presumably due to the narrower redshift bins of the updated sample (mainly due to the inclusion of quasars in the photo-$z$ training that reduced their photo-$z$ errors) and the zero point correction at $\mathrm{Dec}<-29.25^{\circ}$. The shifts are nearly 1-sigma and a reanalysis using updated CMB lensing maps and newer models is in progress. We will present the results in a future publication.

\section{Photometric redshifts}
\label{app:photoz}

Data from the following surveys were used for photo-$z$ training in \cite{Zhou++20} and is reused for the new photo-$z$'s presented here: 2dFLenS \cite{blake_2-degree_2016}, AGES \cite{kochanek_ages_2012}, COSMOS2015 photo-$z$'s \cite{laigle_cosmos2015_2016}, DEEP2 \cite{newman_deep2_2013} and DEEP3 \cite{zhou_deep_2019}, GAMA\cite{baldry_galaxy_2018}, OzDES \cite{lidman_ozdes_2020} (updated to DR2), SDSS Main Galaxy Sample \cite{strauss_spectroscopic_2002} and LRG sample \cite{eisenstein_spectroscopic_2001}, BOSS \cite{dawson_baryon_2013,dawson_sdss-iv_2016} (updated to SDSS DR16 \cite{ahumada_sixteenth_2020}), VIPERS \cite{scodeggio_vimos_2018}, VVDS \cite{le_fevre_vimos_2013} and WiggleZ \cite{parkinson_wigglez_2012}. Quality cuts on these datasets are described in \cite{Zhou++20}.

A major update in the new photo-$z$'s is the inclusion of a large number of additional spectroscopic redshifts for photo-$z$ training, some of which only became available recently: C3R2 DR2 \cite{masters_complete_2019}; eBOSS LRG \cite{ross_completed_2020} and ELG \cite{raichoor_completed_2020} redshifts from the eBOSS LSS catalogs (instead of the SDSS specObj catalog which is based on a different redshifting algorithm); SDSS DR16 quasar catalog \cite{lyke_sloan_2020}; and DESI Survey Validation (SV1) and 1\% Survey (SV3). The DESI data used for photo-$z$ training includes BGS \cite{hahn_desi_2023}, LRG \cite{desi_lrg_paper} and QSO \cite{chaussidon_target_2023} redshifts from SV1 and SV3 and ELG \cite{raichoor_target_2023} redshifts from SV1. We also include redshifts from a special (secondary) DESI program in COSMOS that observed a magnitude-limited sample of galaxies with the same $z$-band fiber-magnitude limit as the DESI LRGs and another magnitude-limited sample of $z>0.7$ galaxies that extends 0.8 magnitudes fainter; this program and the associated data will be described in future work. Except for this DESI special program, all the DESI spectroscopic redshifts used in the photo-$z$ training are publicly available in the DESI Early Data Release, and they are also included in the photo-$z$ catalogs.

The photo-$z$'s are computed with the random forest regression method implemented in SCIKIT-LEARN \cite{pedregosa_scikitlearn_2011}. The following features are used for the random forest: $r$-band magnitude and fiber-magnitude, $g-r$, $r-z$, $z-W1$ and $W1-W2$ colors, half-light radius, aspect ratio (ratio between semiminor and semimajor axes), and a shape parameter that quantifies if a galaxy is better fitted by an exponential profile or a de Vaucouleurs profile (see section 3.1 of \cite{Zhou++20}). Note that the $r$-band fiber-magnitude was not used in the earlier versions of the photo-$z$'s. All the magnitudes and colors are extinction corrected using the Galactic transmission values from LS DR9.

Photo-$z$'s for galaxies at very low redshifts ($z\lesssim0.05$) are often biased high due to the lack of training data (which results from the small volume) at such low redshifts. To mitigate this, we up-weight the low-redshift galaxies by applying the following weights as a function of redshift: $w = 1/(z+0.01)^2+1$ and clipped to constant values at $z<0.0056$ and $z>0.06$. There are similar biases at the high-redshift ($z\gtrsim1$) end (where the photo-$z$'s are biased low), but unlike in \cite{Zhou++20}, we do not up-weight galaxies at the high-redshift end due to issues with up-weighting them when the new training data is added.

The photo-$z$'s are computed with 10-fold cross-validation. Specifically, the training data is randomly divided into 10 subsets, and the photo-$z$'s of each subset is computed using the training data of the combined 9 other subsets. Objects that are not in the training data are randomly assigned to one of the ten training data combinations. This avoids over-fitting for objects in the training data and produces photo-$z$'s with consistent error properties regardless of whether an object is in the training data. The cross-validation also simplifies the assessment of the photo-$z$ performance.

As \cite{Zhou++20}, we do not provide star-galaxy separation estimates. Users of the photo-$z$'s might need to apply cuts to reject stars, e.g.\ by requiring a non-``PSF'' morphological classification or by applying color cuts such as those applied in the DESI LRG selection.

\begin{figure}
    \centering
    \resizebox{0.495\columnwidth}{!}{\includegraphics{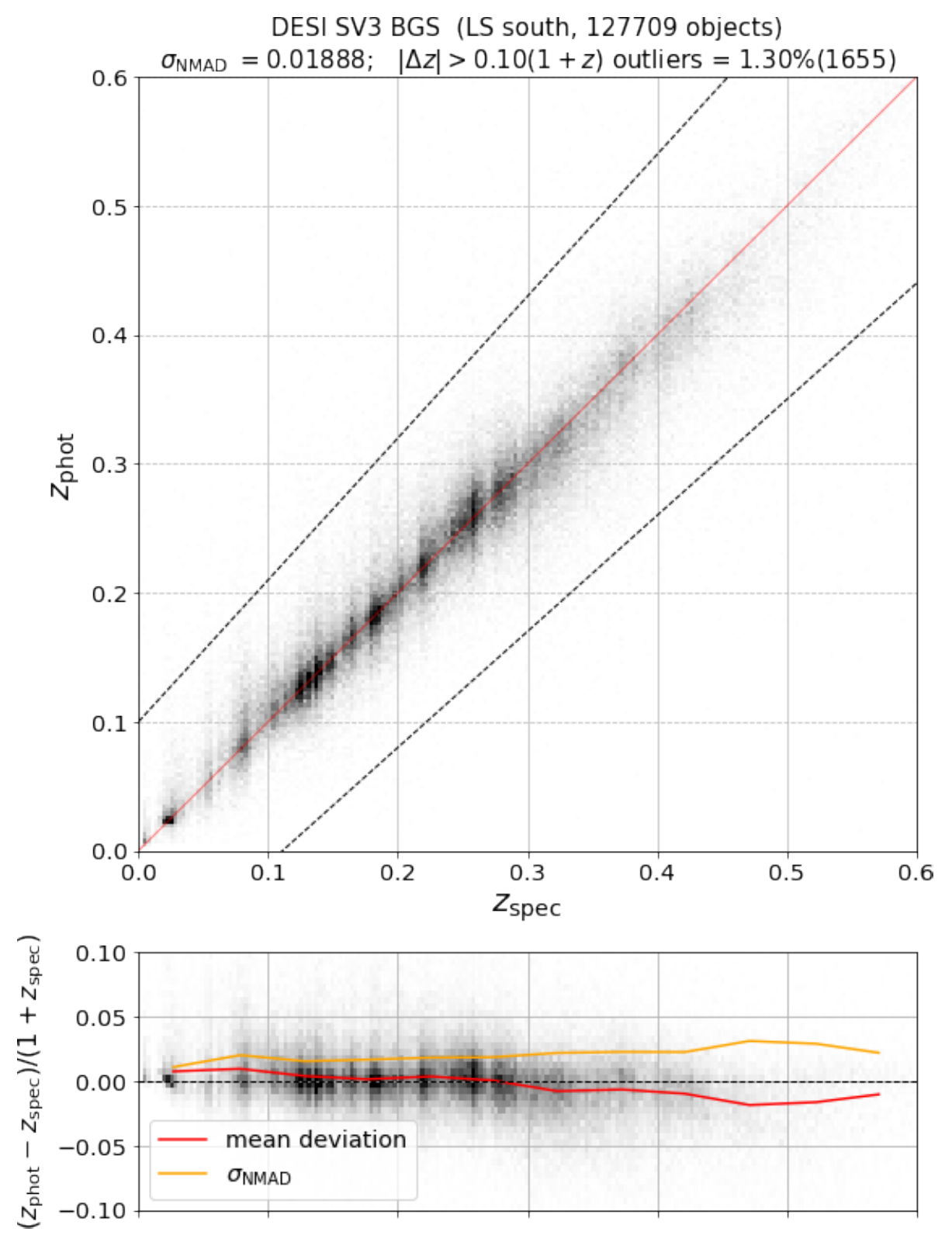}}
    \resizebox{0.495\columnwidth}{!}{\includegraphics{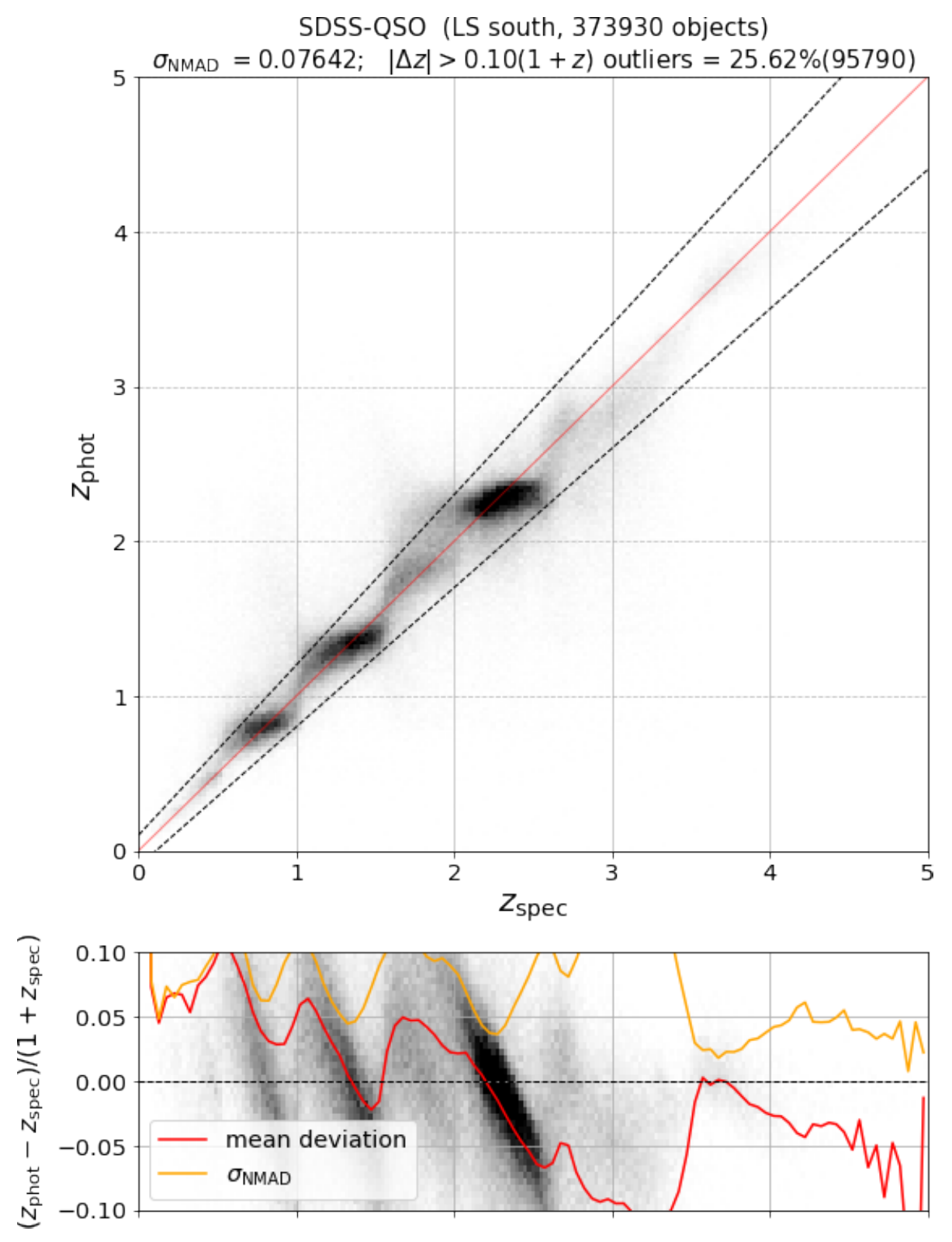}}
    \caption{Photo-$z$ vs spec-$z$ for DESI SV3 BGS sample (left panel) and SDSS DR16 Quasar sample (right panel).}
    \label{fig:more_pz_vs_sz}
\end{figure}

Table \ref{tab:pz_table} lists the training datasets and their photo-$z$ performance. To demonstrate the photo-$z$ performance for non-LRG samples, we show in figure \ref{fig:more_pz_vs_sz} photo-$z$ vs spec-$z$ for the DESI SV3 BGS sample (which has $<2\%$ photo-$z$ scatter) and the SDSS QSO sample. More photo-$z$ vs spec-$z$ plots for all photo-$z$ training datasets can be found on the Legacy Surveys website.

\begin{table}
\centering
\resizebox{1.05\textwidth}{!}{%
\begin{tabular}{c|c|c|c|c|c}
      Survey &         \# objects &              $\sigma_{\mathrm{NMAD}}$ &          $\sigma_{\mathrm{NMAD}}$ ($z_\mathrm{mag}<21$) &       $\eta$ (\%) &  $\eta$ ($z_\mathrm{mag}<21$) (\%) \\
\hline
           2dFLenS &  28573      (0)  &   0.014     (-)   &    0.014     (-)   &     1.4    (-)     &    1.3    (-) \\
              AGES &   4432  (17816)  &   0.022 (0.023)   &    0.022 (0.022)   &     4.0 ( 3.3)     &    3.1 ( 2.3) \\
              BOSS & 967556 (383179)  &   0.014 (0.015)   &    0.014 (0.015)   &     1.0 ( 0.8)     &    0.9 ( 0.7) \\
              C3R2 &   2551    (174)  &   0.101 (0.159)   &    0.026 (0.046)   &    38.1 (52.3)     &    5.0 (12.5) \\
        COSMOS2015 &  64426      (0)  &   0.072     (-)   &    0.032     (-)   &    25.4    (-)     &    5.5    (-) \\
DEEP2 Fields 2,3,4 &  21188  (14835)  &   0.069 (0.079)   &    0.029 (0.032)   &    21.3 (25.5)     &    3.5 ( 4.8) \\
   DEEP2+3 Field 1 &      0  (13480)  &       - (0.086)   &        - (0.032)   &       - (32.2)     &      - ( 5.0) \\
         eBOSS-ELG & 184078   (8780)  &   0.033 (0.041)   &    0.030 (0.036)   &     4.9 ( 7.9)     &    3.2 ( 4.9) \\
         eBOSS-LRG & 107677 (109072)  &   0.020 (0.023)   &    0.020 (0.023)   &     0.5 ( 0.6)     &    0.5 ( 0.5) \\
              GAMA & 127642      (0)  &   0.017     (-)   &    0.017     (-)   &     1.2    (-)     &    1.1    (-) \\
             OzDES &  13062    (894)  &   0.023 (0.026)   &    0.020 (0.022)   &     5.1 ( 7.5)     &    2.2 ( 2.4) \\
              SDSS & 561448 (276299)  &   0.010 (0.010)   &    0.010 (0.010)   &     0.5 ( 0.5)     &    0.4 ( 0.4) \\
          SDSS-QSO & 373930 (283553)  &   0.076 (0.080)   &    0.072 (0.076)   &    25.6 (26.2)     &   23.5 (24.1) \\
            VIPERS &  48262      (0)  &   0.031     (-)   &    0.023     (-)   &     4.7    (-)     &    1.6    (-) \\
              VVDS &   6705      (0)  &   0.050     (-)   &    0.029     (-)   &    18.9    (-)     &    4.9    (-) \\
           WiggleZ & 145850   (1629)  &   0.053 (0.071)   &    0.038 (0.052)   &    18.5 (24.4)     &   10.1 (12.9) \\
       DESI COSMOS &  33231      (0)  &   0.033     (-)   &    0.031     (-)   &     5.5    (-)     &    4.8    (-) \\
      DESI SV1 BGS &  48509  (34688)  &   0.021 (0.023)   &    0.021 (0.023)   &     2.0 ( 2.2)     &    2.0 ( 2.1) \\
      DESI SV1 ELG &  30176  (17735)  &   0.095 (0.118)   &    0.041 (0.055)   &    34.7 (41.2)     &   14.4 (18.7) \\
      DESI SV1 LRG &  16865  (13783)  &   0.028 (0.032)   &    0.025 (0.027)   &     3.3 ( 4.1)     &    2.7 ( 3.0) \\
      DESI SV1 QSO &  12357   (7836)  &   0.103 (0.106)   &    0.080 (0.083)   &    36.9 (37.6)     &   26.6 (27.4) \\
      DESI SV3 BGS & 127709 (113055)  &   0.019 (0.020)   &    0.019 (0.020)   &     1.3 ( 1.4)     &    1.3 ( 1.4) \\
      DESI SV3 LRG &  59855  (55390)  &   0.025 (0.026)   &    0.024 (0.025)   &     1.9 ( 2.1)     &    1.6 ( 1.8) \\
      DESI SV3 QSO &  15498  (14852)  &   0.093 (0.091)   &    0.078 (0.075)   &    32.1 (31.4)     &   25.6 (24.4) \\
\end{tabular}
}
\caption{\label{tab:pz_table} The number of objects, photo-$z$ scatter $\sigma_{\mathrm{NMAD}}$ and outlier fraction $\eta$ (defined in section \ref{sec:photoz}) for each training dataset. We also include the photo-$z$ scatter and outlier fraction for the brighter subset of objects that have $z_\mathrm{mag}<21$. The numbers without parentheses are for the South, and with parentheses for the North. The photo-$z$'s are based on 10-fold cross-validation. 
The number of objects includes duplicates with other surveys (which are removed when the training datasets are combined).}
\end{table}

\section{Modeling the change in fiber-magnitude for a magnified galaxy}
\label{app:fibermagnitude}

Here we describe how we obtain the change in fiber-magnitude $\delta m_\mathrm{fiber}$ for a given change in magnitude $\delta m$ arising from magnification. The $\delta m_\mathrm{fiber}$ value is necessary for properly computing the magnification number count slope.

The DR9 imaging catalog has 5 types of morphology models: point source, round exponential (REX), exponential, de Vaucouleurs and Sersic. Table \ref{tab:morph_type} lists the fraction of LRG targets in each type. For point sources, the fraction of flux within the fiber does not change with magnification, and therefore have $\delta m_\mathrm{fiber} = \delta m_\mathrm{total}$. For extended sources, $\delta m_\mathrm{fiber}$ can be calculated either by numerically integrating the model flux within the fiber radius for the magnified galaxy, or it can be derived from the catalog as long as the galaxy shape parameters are well sampled (which is true for the LS catalog that contains billions of sources). We take the latter approach.

We first obtain the fiber-flux to total flux ratio $f_\mathrm{fiber}/f_\mathrm{total}$ as a function of half light radius and aspect ratio. Since both $f_\mathrm{fiber}$ and $f_\mathrm{total}$ values are available for all sources in the tractor catalog, we obtain $f_\mathrm{fiber}/f_\mathrm{total}$ using the tractor catalog and interpolate between the shape parameters. If a galaxy is magnified, the observed total flux increases by a small fraction of $\epsilon_\mathrm{total}$, and the fiber-flux increases by a different fraction $\epsilon_\mathrm{fiber}$. (Since $\delta m$ and $\epsilon$ are equivalent, hereafter we use $\epsilon$ for simplicity.) Given $f_\mathrm{fiber}/f_\mathrm{total}$ as a function of the shape parameters, we can compute $\epsilon_\mathrm{fiber}/\epsilon_\mathrm{total}$ via finite difference. Below we describe how $f_\mathrm{fiber}/f_\mathrm{total}$ and $\epsilon_\mathrm{fiber}/\epsilon_\mathrm{total}$ are obtained for each morphology model.

\begin{table}
\centering
\begin{tabular}{|r|r|r|}
\hline
Type & All & Faintest 0.01 mag \\
\hline
Point source & 2.2\% & 5.0\% \\
Round exponential & 29.1\% & 52.8\% \\
Exponential & 9.8\% & 13.6\% \\
de Vaucouleurs & 42.9\% & 27.5\% \\
Sersic & 16.0\% & 1.0\% \\
\hline
\end{tabular}
\caption{\label{tab:morph_type} Morphological type composition of all Main LRG targets. Since our sensitivity to magnification is mostly determined by the faintest objects (although brighter sources also matter due to the photo-$z$ binning), we also list the composition of the LRGs within 0.01 magnitude of the $z$-band fiber-magnitude cut. We use all LRGs when computing the number count slope via finite difference.}
\end{table}

\begin{figure}
    \centering
    \resizebox{0.7\columnwidth}{!}{\includegraphics{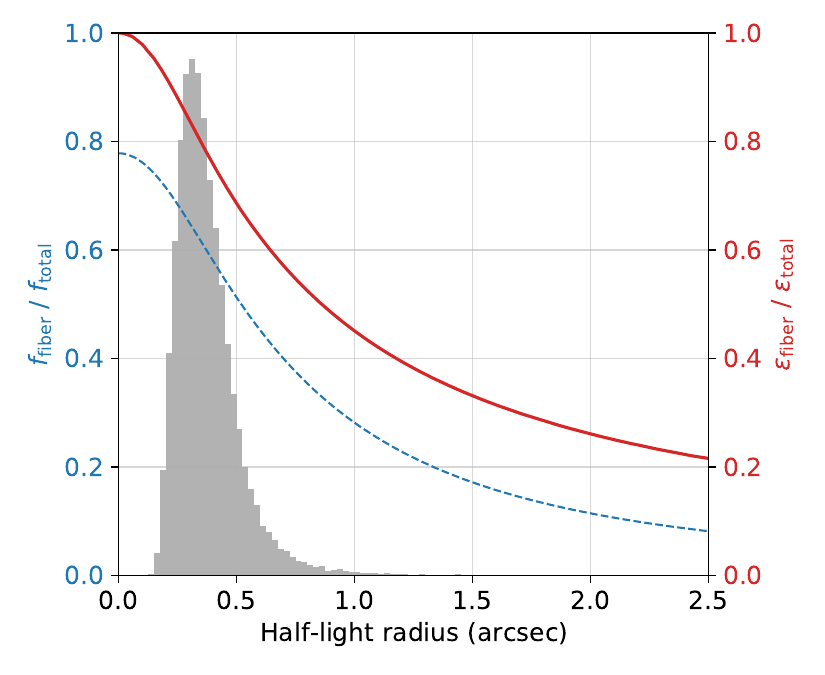}}
    \caption{The fiber-flux to total flux ratio ($f_\mathrm{fiber}/f_\mathrm{total}$, blue dashed curve) and the ratio between the fractional changes of the two fluxes ($\epsilon_\mathrm{fiber}/\epsilon_\mathrm{total}$, red solid curve) as a function of half-light radius, for round exponential galaxies. The $\epsilon_\mathrm{fiber}/\epsilon_\mathrm{total}$ values are used in calculating the ``magnified'' fiber-fluxes. The distribution of the half-light radii for round exponential LRGs within 0.01 magnitude of the $z$-band fiber-magnitude cut is shown in the gray histogram.}
    \label{fig:rex_fiberflux}
\end{figure}

Round exponential galaxies, which constitute most of the LRG targets near the fiber-flux cut, have only one shape parameter, the half-light radius. We measure the mean $f_\mathrm{fiber}/f_\mathrm{total}$ in bins of half-light radius, and interpolate with a cubic spline. We then use the interpolated function to obtain $\epsilon_\mathrm{fiber}/\epsilon_\mathrm{total}$ with finite difference. The measured $\epsilon_\mathrm{fiber}/\epsilon_\mathrm{total}$ is noisy at radius above 2 arcsec, so we use the Savitzky–Golay filter to obtain a smoothed function. Figure \ref{fig:rex_fiberflux} shows $f_\mathrm{fiber}/f_\mathrm{total}$ and the smoothed $\epsilon_\mathrm{fiber}/\epsilon_\mathrm{total}$ as a function of half-light radius.

\begin{figure}
    \centering
    \resizebox{1.0\columnwidth}{!}{\includegraphics{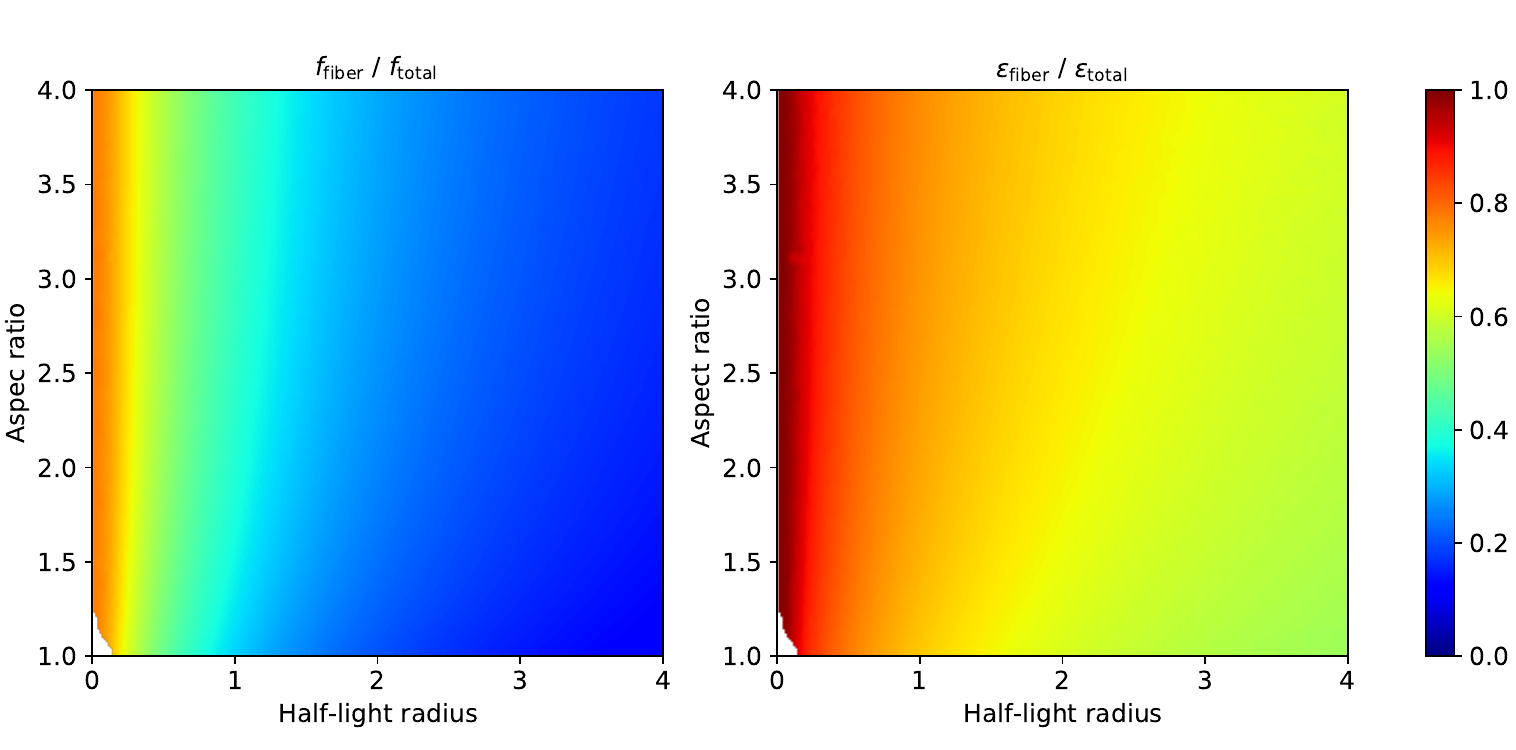}}
    \caption{Similar to figure \ref{fig:rex_fiberflux}, but for de Vaucouleurs galaxies with an additional shape parameter, the aspect ratio.}
    \label{fig:dev_fiberflux}
\end{figure}

Exponential and de Vaucouleurs galaxies have two shape parameters, half-light radius and aspect ratio (the ratio of the major axis to the minor axis). The modeling is similar to round exponential galaxies, except that the binning and interpolation are done in 2-D instead of 1-D. Figure \ref{fig:dev_fiberflux} shows $f_\mathrm{fiber}/f_\mathrm{total}$ and the smoothed $\epsilon_\mathrm{fiber}/\epsilon_\mathrm{total}$ as a function of half-light radius and the aspect ratio for de Vaucouleurs galaxies.

Sersic galaxies have an additional parameter, the Sersic index. Since only $1\%$ of the LRG targets near the fiber-flux cut are Sersic, instead of attempting to create a 3-parameter model, we simply approximate the Sersic galaxies with either exponential or de Vaucouleurs profile, depending on the Sersic index $n$: exponential for $n<2.5$ and de Vaucouleurs for $n\geq2.5$.

\bibliographystyle{JHEP}
\bibliography{main}
\end{document}